\documentclass
    [   aip, jcp
	, superscriptaddress
	, showpacs
	, floatfix
        , citeautoscript 
	  , twocolumn , reprint
    ]
    {revtex4-1}

\usepackage{graphicx}
\usepackage{epsfig,amsmath, amsfonts, amssymb, xcolor}
\usepackage{stmaryrd,skak,marvosym} 
\usepackage{soul}
\usepackage{textcomp}
\usepackage{hyperref}

\newcommand{\of}[1]{\ensuremath{\left(#1\right)}}
\newcommand{\abs}[1]{\ensuremath{\left\vert#1\right\vert}}
\newcommand{\fd}[1]{\ensuremath{\left[#1\right]}}
\newcommand{\set}[1]{\ensuremath{\left\lbrace#1\right\rbrace}}

\newcommand{\thermal}[1]{\ensuremath{\left\langle#1\right\rangle}}

\newcommand{\cur}[1][]{\ensuremath{\mathcal{C}_{#1}}}

\newcommand{\tcb}[1][]{\ensuremath{T^{\of{s}}_{c}}}
\newcommand{\Op}[1][]{\ensuremath{\phi_{#1}}}
\newcommand{\T}[1][]{\ensuremath{\mathcal{T}_{#1}}}
\newcommand{\ccf}[1][]{\ensuremath{f_{C#1}}}
\newcommand{\ccp}[1][]{\ensuremath{U_{c#1}}}
\newcommand{\sfeos}[1][]{\ensuremath{F_{#1}}}
\newcommand{\sff}[1][]{\ensuremath{\vartheta_{#1}}}
\newcommand{\sfp}[1][]{\ensuremath{\theta^{#1}}}

\newcommand{\Ham}[1][]{\ensuremath{\mathcal{H}}}
\newcommand{\hb}[1][]{\ensuremath{h_{b#1}}}
\newcommand{\hs}[1][]{\ensuremath{h_s}}
\renewcommand{\c}[1][]{\ensuremath{c}}

\newcommand{\Fe}[1][]{\ensuremath{\mathcal{F}_{#1}}}
\newcommand{\fe}[1][]{\ensuremath{{\mathfrak{f}_{#1}}}}

\newcommand{\hSt}[1][]{\ensuremath{\mathfrak{h}_{#1}}}

\newcommand {\dv}[1][]{\ensuremath{\partial{}v}}

\newcommand{\bsec}[1][]{\ensuremath{B_{2#1}}}
\newcommand{\w}[1][]{\ensuremath{W}}

\newcommand{\m}[1][]{\ensuremath{\omega_{#1}}}

\newcommand{\x}[1][]{\ensuremath{\mathbf{r}_{#1}}}
\newcommand{\q}[1][]{\ensuremath{\mathbf{q}_{#1}}}
\newcommand{\pot}[1][]{\ensuremath{U_{#1}}}

\newcommand{\sgn}[1][]{\ensuremath{\mathop{sgn}}}
\newcommand{\ft}[1]{\ensuremath{\hat{#1}}}

\newcommand{\eref}[1]{{Eq.~\eqref{#1}}}
\newcommand{\fref}[1]{{Fig.~\ref{#1}}}

\begin{document}
\title{Phase behavior of colloidal suspensions with 
    critical solvents in terms of effective interactions}
\date{\today}
\author{T. F. Mohry}
\email{mohry@is.mpg.de}
\affiliation{Max-Planck-Institut f{\"u}r Intelligente Systeme,
  Heisenbergstra{\ss}e 3, 70569 Stuttgart, Germany}
\affiliation{Universit{\"a}t Stuttgart, Institut f{\"u}r Theoretische und Angewandte Physik,
  Pfaffenwaldring 57, 70569 Stuttgart, Germany} 
\author{A. Macio{\l}ek}
\email{maciolek@is.mpg.de}
\affiliation{Max-Planck-Institut f{\"u}r Intelligente Systeme,
  Heisenbergstra{\ss}e 3, 70569 Stuttgart, Germany}
\affiliation{Universit{\"a}t Stuttgart, Institut f{\"u}r Theoretische und Angewandte Physik,
  Pfaffenwaldring 57, 70569 Stuttgart, Germany}
\affiliation{Institute of Physical Chemistry, Polish Academy of Sciences,
  Kasprzaka 44/52, PL-01-224 Warsaw, Poland}
\author{S. Dietrich}
\email{dietrich@is.mpg.de}
\affiliation{Max-Planck-Institut f{\"u}r Intelligente Systeme,
  Heisenbergstra{\ss}e 3, 70569 Stuttgart, Germany}
\affiliation{Universit{\"a}t Stuttgart, Institut f{\"u}r Theoretische und Angewandte Physik,
  Pfaffenwaldring 57, 70569 Stuttgart, Germany}
\begin{abstract}
	We study the phase behavior of colloidal suspensions 
	the solvents of which are considered to be binary liquid 
	mixtures undergoing phase segregation. We focus on the thermodynamic 
	region close to the critical point of the 
	accompanying miscibility gap.  There, due  to the
	colloidal particles acting as cavities in the critical medium, 
	the spatial confinements of the critical fluctuations of the corresponding 
	order parameter result in the effective, so-called 
	critical Casimir forces between the colloids. Employing an approach 
	in terms of effective, {\it one}-component colloidal systems,  
	we explore the possibility of phase coexistence between 
	two phases of colloidal suspensions,  one being rich and 
	the other being poor in colloidal particles. The 
	reliability of this effective approach is discussed.
\end{abstract}
\maketitle
\section{Introduction}
In colloidal suspensions the dissolved particles,  typically micrometer-sized, 
interact via  effective interactions which  are a combination of  direct
interactions, such as electrostatic or van der Waals interactions,
and   indirect, effective ones due to the presence of
other, smaller solute particles or mediated by the solvent \cite{Likos:2001}.
For example, by adding   non-adsorbing polymers to the solution
depletion interactions  between  colloidal particles can be induced which
are predominantly attractive \cite{Asakura-et:1954}. The range and the strength
of this entropy-driven  attraction can be  manipulated by varying the polymer-colloid 
size ratio  or  the polymer concentration. The resulting phase diagrams of the
colloids are very sensitive  to changes in the depletion-induced pair potential 
between colloids \cite{Likos:2001,Dijkstra-et:1999}.
It is also possible to induce such attractive depletion forces by adding much
smaller colloidal particles or by using solvents with surfactants which form
micelles acting as depletion agents \cite{Buzzaccaro-et:2007}.

In contrast to depletion forces, solvent mediated
interactions between colloidal particles can depend sensitively on the 
{\it thermodynamic state} of the solvent \cite{Evans:1990}.
This is the case if the solvent exhibits fluctuations on
large spatial scales such as the fluctuations near the
surfaces of the colloidal particles associated with wetting phenomena 
near a first-order phase transition of the solvent or the thermal
fluctuations of the solvent order parameter near a second-order
phase transition. 
In the latter case, critical fluctuations are correlated over distances 
proportional to the correlation length $\xi$ which diverges upon 
approaching the critical point of the unconfined
{\it{}s}olvent (without colloids) at the critical temperature \tcb{}.
Exposing the near-critical fluid to boundaries, e.g., by inserting
colloidal particles acting as cavities, perturbs the fluctuating
order parameter near the surfaces of the colloids on
the scale of $\xi$ and restricts the spectrum
of its thermal fluctuations. Since these restrictions depend on the spatial
configuration of the colloids, they result \cite{Fisher-et1978} 
in an effective  force between the particles 
which is called the critical Casimir force (CCF)  $f_C$.
Accordingly the range of the CCF
is proportional to the bulk correlation length $\xi$.
Therefore it can be tuned {\it continuously}  by small changes of  the
temperature $T$. The range  can also be   controlled by varying the conjugate
ordering field \hb{} of the order parameter of the solvent such as
the chemical potential in the case of a simple fluid  or the chemical potential 
difference of the two species forming a  binary liquid mixture.
The strength and the sign of $f_C$ can be manipulated as well.
This can be achieved  by  varying the temperature  $T$
or \hb{} and  by suitable surface treatments, respectively
\cite{Hertlein-et2008,Gambassi-et:2009,Nellen-et:2009}.
Compared with other effective forces between colloid particles CCFs offer two
distinct features. First, due to the universality of critical phenomena, to
a large extent CCFs do not depend on the microscopic details of the system.
Second, whereas adding depletion agents or ions changes effective forces
irreversibly, the tuning of $f_C$ via $T$ is fully and easily reversible.

Although the CCFs between two colloidal particles depend on  the
(instantaneous) spatial configuration of {\it all} colloids, one can
consider  dilute  suspensions or temperatures sufficiently far away
from \tcb{}, such that
the range $2R+\xi$ of the critical Casimir interaction between the
colloids of radius $R$ is much smaller than
the mean distance between them.
For these cases the assumption of pairwise additive CCFs is  expected
to be valid. Using this pairwise approximation we are able to map
the actual system of a mixture of (monodisperse) colloidal particles
and  solvent molecules to an
{\it  effective, one-component} system of colloidal
particles, in which  the presence and influence of the other
components of the mixture  enter  through the parameters of
the effective pair potential.
Adopting this approach allows us to use standard liquid state theory 
in order to determine the structure of an ensemble of colloidal particles 
immersed in a near-critical solvent and to study its sensitivity 
to changes of the critical Casimir potential 
due to temperature variations.
Since within this effective approach the  feedback of the 
colloids on the solvent and its critical behavior is neglected,
this way the phase behavior of the full
many-component  system cannot be determined in all details.
However, for those parameters of the
thermodynamic phase space for which this 
approach is  applicable, we predict a colloidal ``liquid''-``gas'' 
phase coexistence, i.e., the coexistence of two phases which differ 
with respect to their colloidal number densities.

The necessary input for the approach employed in the present study
is the CCF  $f_C$  between two spherical particles.
It is known that at the bulk critical point $f_C$ is long-ranged.
In the so-called protein limit corresponding to $D/R$, $\xi/D \gg 1$,
where $R$ is the colloid radius and $D$ the
surface-to-surface distance between the colloids,
the so-called small-sphere expansion \cite{colloids2} renders 
$f_C(D;\tcb,R)/\of{k_B\tcb} \sim R^{d-2+\eta}D^{-(d-1+\eta)}$,
where $d$ is the spatial dimension
and $\eta$ is the standard bulk critical exponent for the
two-point correlation function.
In the  opposite, so-called Derjaguin limit $D \ll R$ 
one has \cite{colloids2}
$f_C(D;\tcb,R)/\of{k_B\tcb} \sim R^{ \of{d-1}/2}D^{-\of{d+1}/2}$.
Thus the CCF  can indeed successfully compete with direct dispersion
\cite{dantchev_dietrich} or electrostatic forces in determining the
stability and phase behavior of colloidal systems.
Away from  the critical point, according to finite-size  scaling theory
(see, e.g., Ref.~\onlinecite{FSS}), the CCF exhibits  scaling
described by a universal  scaling function  which
is determined solely by  the so-called universality class (UC)  of the
phase transition occurring in the
bulk, the geometry, and the surface 
universality classes of the confining surfaces
\cite{Diehl:1986,Krech:1990:0,Dbook,gambassi}.
The relevant UC for the present study is the Ising UC with
symmetry-breaking boundary conditions. For spherical particles,
theoretical predictions for the universal scaling function
$\vartheta$ of the CCF
in the full range of parameters are available only within
mean-field theory (MFT)   \cite{colloids1a,colloids1b}
(for $\hb=0$ and ellipsoidal particles see Ref.~\onlinecite{khd-08}).
In contrast, for planar surfaces in $d=3$ results beyond MFT are available.
For a vanishing bulk ordering field $\hb=0$
and symmetry breaking surface fields 
they are provided by field-theoretical studies \cite{krech},
the extended de Gennes-Fisher local-functional
method \cite{upton,FdeG_loc_fun}, 
and Monte  Carlo (MC) simulations  
 \cite{krech,vas,Hasenbusch,Hasenbusch-cross,vas-cross}. 
Moreover, in Ref.~\onlinecite{Buzzaccaro-et:2010} results for the 
scaling function of the CCF for $\hb\ne 0$ are presented which 
are based on a density functional approach.
Also corresponding experimental data
 \cite{Hertlein-et2008,Nellen-et:2009,Gambassi-et:2009,pershan,rafai}
are available.
Based on the Derjaguin approximation  \cite{Derjaguin:1934} the 
knowledge of the scaling function of the  CCF acting between two
parallel surfaces can be used to obtain the scaling function for $f_C$
between two spheres and between a sphere and a planar wall
\cite{colloids1a,colloids1b}. The Derjaguin  approximation assumes
that  the surface-to-surface distance $D$ between two spheres is much
smaller than their radius  $R$.
In many cases, however, this approximation works surprisingly well 
\cite{Gambassi-et:2009,Troendle-et:2010} even for $D\lesssim R$.
We shall use this approximation for temperatures
which  correspond to $\xi \lesssim R$, because under
this condition  the CCFs between the colloids act only at
surface-to-surface distances $D$ small compared with  $R$.
In order to handle the dependence of the CCF on \hb{}
within the Derjaguin approximation, we propose a
suitable approximation for the film scaling
function of the CCF in $d=3$. A necessary input for
this latter approximation is the mean-field scaling
function for the CCF, which we have
calculated using the field-theoretical approach.

Experimental studies of the phase behavior of colloidal suspensions
with phase separating solvents have encompassed silica spheres
immersed in water-2-butoxyethanol \of{C_4E_1} mixtures and in 
water-lutidine mixtures \cite{Kline-et:1994,Jayalakshmi-et:1997}. 
In Ref.~\onlinecite{Koehler-et:1997} both silica and polystyrene 
particles immersed in these binary mixtures have been studied 
focusing on the formation of colloidal crystals and its relation to
aggregation phenomena. A few theoretical \cite{Sluckin:1990} and
simulation \cite{Loewen:1995,Netz:1996} attempts
have been concerned with such kinds of colloidal suspensions.

Our paper is organized such that in Sect.~\ref{sec:theory} we summarize 
the theoretical background of our analysis.
In Subsect.~\ref{sec:colloids} we discuss colloidal suspensions and the
effective model we use, whereas Subsect.~\ref{ssc:critphen} provides
information concerning critical phenomena and the CCF. 
For the parameters entering into the effective potential between the
colloids, in Subsect.~\ref{ssc:res_range} we discuss the range of  
their values corresponding to possible experimental realizations.
In Subsect.~\ref{ssc:res_thermo} the thermodynamics of the 
considered colloidal suspensions is analyzed.
A general discussion of the phase diagram of the 
actual ternary mixtures consisting of the colloids and the binary
solvent is given in Subsects.~\ref{subsec:td_generals} and \ref{ssc:sc}. 
The reliability and the limitations of the effective, one-component
approach are discussed in Subsect.~\ref{ssc:eff}. The 
results for the phase diagrams emerging from
the effective approach are described in Subsect.~\ref{ssc:pd}.
In Sect.~\ref{sec:summary} we conclude with a summary.

\section{Theory}
\label{sec:theory}

\subsection{Colloidal suspensions \label{sec:colloids}}

\subsubsection{Interactions \label{ssc:coll_int}}

The effective interactions between colloidal particles are rich, subtle,
and specific due to the diversity of materials and solvents which can
be used. Our goal is to provide a general view  of the effects a critical
solvent has on dissolved colloids due to the  emerging {\it universal} CCFs.
Therefore we adopt a background interaction potential between the colloids
which is present also away from \tcb{} and captures only the essential
features of a stable suspension on the relevant, i.e., mesoscopic, length scale.
These features are the hard core repulsion for center-to-center
distances $r<2R$ and a  soft, repulsive  contribution
\begin{equation}
\begin{split}
    \label{eq:coll_repulsion}
    V_{rep}\of{r}/(k_BT)=&\,\pot[rep]\of{r}  \\
		=&\,A \exp\of{-\kappa D},
		\,\; D=r-2R>0,
\end{split}
\end{equation}
which prevents coagulation favored by effectively
attractive dispersion forces. The main  mechanisms
providing  $\pot[rep]\of{r}$ are either electrostatic or
steric repulsion, which are both described by the generic functional
form given by  \eref{eq:coll_repulsion} \cite{Barrat-et:2003,Russel-et:1989}.
The steric repulsion is achieved by a polymer coverage of the colloidal surface.
If two such covered colloids come close to each other the polymer layers overlap, which
leads to a decrease in their configurational entropy and thus to an effective repulsion.
Concerning the electrostatic repulsion, for large values of the surface-to-surface
distance $D$ the effective interaction between the corresponding electrical double-layers
at the colloid  surfaces dominates; this leads to a repulsion.
The range $\kappa^{-1}$ of the repulsion is associated with the Debye screening length in
the case of electrostatic repulsion  and with the polymer length in the case of the steric
repulsion. The strength $A$ of the repulsion depends on the colloidal surface charge
density and on the polymer density, respectively.
For the effective Coulomb interaction screened by
counterions $A$ is given by \cite{Russel-et:1989}
\begin{equation}
\label{eq:A_electric}
	A= \of{\epsilon \epsilon_0 }^{-1} \Upsilon^{2}\kappa^{-2}R/\of{k_B T},
\end{equation}
where $\epsilon$ is the permittivity of the solvent relative to 
vacuum, $\epsilon_0$ is the permittivity of the vacuum, $\Upsilon$ 
is the surface charge density of the colloid, and $\kappa$ is 
the inverse Debye screening length.

In our present study we shall analyze the behavior of monodisperse
colloids immersed in a near-critical solvent. We treat the solvent
in an effective way, i.e., we do not consider the full many  component
system but an effective one-component system of colloids  
for which  the presence of the solvent enters via the effective
pair potential $V\of{r=D+2R}/(k_BT)=\pot\of{r=D+2R}$
(see  \eref{eq:coll_repulsion}):
\begin{widetext}
\begin{equation}
\label{eq:coll_potential0}
    \pot\of{r}=
    \begin{cases}
      \infty, & D<0 \\
      \pot[rep]+\pot[c]^{(d=3)}=
      A\exp(-\kappa D)+(1/\Delta)\theta^{(d=3)}(\Theta,\Delta, \Sigma),  & D>0.
    \end{cases}
\end{equation}
\end{widetext}
In \eref{eq:coll_potential0} $\pot[c]^{(d=3)}$ is the critical Casimir
potential and  $\theta^{(d=3)}$  is its universal scaling function
in   spatial dimension  $d=3$. 
For  the  present system,
$\theta^{(d)}$ is a function of the three scaling variables
$\Theta=\sgn\of{t} D/\xi$, $\Delta=D/R$, and 
$\Sigma=\sgn\of{\hb}\xi/\xi^{\of{h}}$.
Here $\xi\of{t\gtrless 0}=\xi_0^{\pm}\abs{t}^{-\nu}$, 
with  $t=\pm(T-\tcb{})/\tcb{}$  for an upper ($+$) 
and a lower ($-$) critical point, respectively,
is the true correlation length governing the exponential decay of the 
solvent  bulk OP correlation function for $t\to0^{\pm}$ and for the
ordering field, conjugated to the OP, $\hb=0$. 
The amplitudes $\xi_0^{\pm}$ ($\pm$ referring to
the sign of $t$) are non-universal but their ratio $\xi_0^+/\xi_0^-$ 
is universal. The correlation length 
$\xi^{\of{h}} =\xi^{\of{h}}_0\abs{\hb}^{-\nu/\of{\beta\delta}}$
governs the exponential decay of the solvent bulk OP
correlation function for $t=0$ and $\hb\to0$, where $\xi^{\of{h}}_0$
is a non-universal amplitude. (Note that $\xi^{\of{h}}_0$
has not the dimension of a length, see Appendix~\ref{app:1}; 
$\nu$, $\beta$, and $\delta$ are standard bulk critical exponents 
\cite{Pelissetto-et:2002}.)
The critical finite-size scaling and the scaling variables will be 
discussed in detail in Subsect.~\ref{ssc:critphen}.

We do not consider an additional interaction which would account for 
effectively attractive dispersion forces. Effectively, dispersion forces 
can be switched off by using index-matched colloidal suspensions.
As will be discussed in Sect.~\ref{sec:results}, 
the presence of attractive dispersion forces does not change the 
conclusions of the present study. For a particular experimental 
realization $A$, $\kappa$, and $R$ 
are material dependent constants.

As mentioned in the Introduction, the scaling function
$\theta^{(d)}$ of the critical Casimir potential between two
spheres is not known beyond MFT for the full range of the
scaling variables $\Theta$, $\Delta$, and $\Sigma$.
In order to overcome this restriction we shall use 
two approximations for $\theta^{(d=3)}$. 
First, we use the Derjaguin approximation
(c.f., \eref{eq:sfpot_derjaguin}) in order to express 
$\theta^{(d)}$ in terms of the universal scaling function
$\vartheta_{\parallel}^{\of{d}}$  of the CCF  for the film
geometry. The latter is known rather accurately from
 MC   simulations  in $d=3$
and for $\hb=0$; we shall use this knowledge in the present study.
 Second,  in order to be able to capture the
dependence of CCFs on $\hb$ we propose an approximation for
$\vartheta_{\parallel}^{\of{d}}$ as a function of $\Sigma$.
Our approximation is constructed in such a way that for $\hb\to 0$
the scaling function reduces  exactly to
$\vartheta_{\parallel}^{\of{d}}\of{\Sigma=0}$
for all $d$, and at \tcb{} its shape is the same
as within MFT (see, c.f., \eref{eq:sfp_h_approx}).
These two approximations will be discussed extensively
in Subsect.~\ref{ssc:critphen}.

\subsubsection{Thermodynamics and stability \label{ssc:coll_stability}}

Determining the thermodynamic properties of a system from the underlying 
pair potential $V\of{r}$ of its constituents is a central issue of 
statistical physics \cite{Hansen-et:1976}. In principle the thermodynamic
properties of a system can be determined from its 
correlation functions \cite{Hansen-et:1976,Caccamo:1996}.
For example, the so-called virial equation provides the pressure $p$
of the homogeneous system:
\begin{equation}
\label{eq:pressure_oz}
	p/(\rho{}k_B T) =
	1- \frac{2}{3}\pi \rho \int_{0 }^{\infty}
	U'\of{r} g\of{r} r^3 \mbox{ d}r.
\end{equation}
The  radial distribution function $g\of{r}$ is related to the total 
correlation function (TCF) $h\of{r}$ according to $h\of{r}=g\of{r}-1$. 
The isothermal compressibility $\chi_T$ follows from the sum
rule \cite{Hansen-et:1976}
\begin{equation}
\label{eq:compressibility}
    \lim\limits_{q\to 0}S\of{q}= \rho k_B T \chi_T,
\end{equation}
where the structure factor
\begin{equation}
\label{eq:structurefactor}
    S\of{q}=1+ \rho \ft{h}\of{q}= 1/\of{1-\rho \ft{c}\of{q}} 
\end{equation}
can be determined by scattering experiments. 
The system becomes unstable for  $\chi_T\to \infty$, corresponding
to the critical point and, within MFT, to the spinodals in the 
phase diagram (see, e.g., Ref.~\onlinecite{Binder-et:1978}). 
The TCF $h\of{r}$ is related to the direct correlation function (DCF)
$c\of{r}$ and the number density $\rho$ via the 
Orstein-Zernicke equation \cite{Ornstein-et:1914,Hansen-et:1976}: 
$h\of{r}=c\of{r}+\rho\int h\of{r^{\prime}}c\of{\abs{\x-\x^{\prime}}}\text{ d}^{3}\x^{\prime}$. 
The correlation functions can be calculated iteratively 
\cite{Hansen-et:1976,Gillian:1979}. For a given approximate 
expression for $c_{i}\of{r}$ one obtains $h_i\of{r}$ 
in Fourier space:
\begin{equation}
  \label{eq:oz}
  \ft{h}\of{q}= \ft{c}\of{q} /\of{1-\rho \ft{c}\of{q}},
\end{equation}
with
$\ft{h}\of{\q}=\int e^{i\q\x}h\of{\x}\text{ d}^{d}\x$,
analogously for $\ft{c}\of{\q}$, and $q=\abs{\q}$.
By choosing a bridge function $b\of{r}$, the closure
\begin{equation}
	\label{eq:closure}
	h\of{r}+1=\exp \set{ -\pot\of{r} +h\of{r}-c\of{r} +b\of{r} }
\end{equation}
renders a DCF $c_{i+1}$  which typically differs from $c_{i}$. 
This procedure is continued until satisfactory convergence is 
achieved. The initial guess $c_{i=1}$  is guided by the
shape of the direct interaction potential. For the bridge 
function we use the so-called Percus-Yevick approximation (PY), 
\begin{equation}
  \label{eq:py_bridgefct}
  b_{PY}\of{r}= \ln\fd{h\of{r}-c\of{r}+1}-h\of{r}+c\of{r},
\end{equation}
and the so-called hypernetted-chain approximation (HNC), 
\begin{equation}
  \label{eq:hnc_bridgefct}
  b_{HNC}\of{r}= 0.
\end{equation}
In terms of the DCF $c\of{r}$ and of $k\of{r}=h\of{r}-c\of{r}$
(which is more useful than $h\of{r}$ for handling 
the hard core $\pot\of{r<2R}=\infty$) the PY closure 
(\eref{eq:py_bridgefct}) can be written as
\begin{equation}
\label{eq:py_k}
	{c}_{PY}\of{r}= \left[\exp\set{-\pot\of{r}}-1\right]\of{k\of{r}+1}
\end{equation}
whereas the HNC closure (\eref{eq:hnc_bridgefct}) 
leads to
\begin{equation}
\label{eq:hnc_k}
	c_{HNC}\of{r}= \exp\set{-\pot\of{r} +k\of{r}}-k\of{r}-1 .
\end{equation}
For a more detailed discussion of the applicability and reliability
of this integral equation approach  (IEA)  we refer 
to Ref.~\onlinecite{Caccamo:1996}.
For comparison with   MC  simulations in the case of a potential with 
attractive and repulsive parts see for example Ref.~\onlinecite{Archer-et:2007},
which discusses particles interacting with a pair potential containing 
attractive and repulsive  Yukawa-like contributions
$\epsilon_{i} e^{-\kappa_{i} r}/r$. The  IEA is  capable to reveal the 
rich phase behavior of such systems.

There are further relations expressing  thermodynamic 
quantities in terms of correlation functions which are exact
from the formal point of view. However, because
the bridge function is not known exactly, the set
of equations \eqref{eq:oz} and \eqref{eq:closure}  may have no solution
in the full one-phase region of the thermodynamic
phase space; moreover the
resulting thermodynamic quantities depend on the scheme taken. 
This is the well known thermodynamic inconsistency of this approach 
(although there are more sophisticated schemes trying to cope with 
this problem) \cite{Caccamo:1996}.
Moreover, for the kind of systems considered here, the determination 
of phase equilibria  is even more subtle because
due to the adsorption phenomena, which are state
dependent, the effective potential between the colloids
depends on the thermodynamic state itself.
Inter alia this implies that the effective potential acting between 
the particles should be different in coexisting phases.
This feature is not captured by the effective potential approach
presented above.
Therefore within the  IEA only certain estimates
for the coexistence curve can be obtained. For a reliable phase diagram
actually the full many component mixture  (such as
the binary solvent plus the colloidal particles) has to be considered.

Another useful first insight into the collective behavior
of attractive (spherical) particles is provided by the 
second virial coefficient  \cite{Hansen-et:1976}
\begin{equation}
\label{eq:secvirial}
    \bsec=2\pi \int_{0}^{\infty}\of{1-\exp\set{-U\of{r}}}r^{2} \text{ d}r.
\end{equation}
Beyond the ideal gas contribution it determines the leading non-trivial
term in the expansion of the pressure
$p\of{\rho}/\of{k_BT\rho}=1+\bsec{}\rho+\ldots$ in terms of powers of
the number density $\rho$. 
Measurements of \bsec{} for colloids immersed 
in near-critical solvents have been reported 
in Ref.~\onlinecite{Kurnaz}.
Vliegenthart and Lekkerkerker \cite{Vliegenthart-et:2000} and
Noro and Frenkel \cite{Noro-et:2000} (VLNF) proposed an extended law
of corresponding states according to which the value of the reduced 
second virial coefficient
$\bsec^{*}\equiv\bsec/\bsec^{\of{HS}}$
at the critical point is the same for all systems 
composed of particles with short-ranged attractions, regardless of the 
details of these interactions.
$\bsec^{\of{HS}}=\frac{2\pi}{3}\sigma^3$ is the second virial coefficient
of a suitable reference system of hard spheres 
(HS) with diameter $\sigma$.
This (approximate) empirical rule is supported by experimental data
\cite{Vliegenthart-et:2000} and by theoretical results \cite{thN}.
The critical value $\bsec[,c]^{*}$ can be
obtained in particular from the Baxter model for adhesive
hard spheres \cite{Baxter:1968}, for which the interaction is given by
$\exp\set{-\pot\of{r}}=
H \of{r-\sigma}+\frac{\sigma}
{12 {\mathtt S}}\delta\of{r-\sigma}$ 
where $H\of{r}$ is the Heaviside function and $\delta(r)$ is the delta function.
The reduced second virial coefficient is related to the so-called
stickiness parameter ${\mathtt S}$ by
\begin{equation}
\label{eq:stickyparam}
    \bsec^{*}\equiv\bsec/\of{\frac{2\pi}{3}\sigma^3}=1-\frac{1}{4 {\mathtt S}}.
\end{equation}
This model exhibits a liquid-vapor phase transition as function of 
${\mathtt S}$ with the critical value \cite{Miller-et:2003}
${\mathtt S}_c\simeq 0.113$, so that $\bsec[,c]^*\simeq-1.212$.

As  discussed in Subsect.~\ref{ssc:res_range} under
typical experimental conditions the bulk correlation length  $\xi$ 
and therefore the range of the CCFs are  smaller than the radius $R$  
of the (micron-sized) colloids.
Therefore the main attractive contribution to the resulting effective  
pair potential between the colloidal particles is localized 
at a range of distances $D$  which are small compared with $R$.
Therefore for experimentally realizable conditions the effective pair
potential can be considered to be short-ranged and thus matches
the conditions described above for the approximate VLNF conjecture. 
Accordingly, close to the corresponding value of $\bsec[,c]^*$
the effectively one-component system  of colloidal particles can be 
expected to exhibit a critical point terminating
a ``liquid''-``gas'' phase separation line. 

In order to obtain a suitable HS reference system,  following
Weeks, Chandler, and Andersen  \cite{Weeks-et:1971,Andersen-et:1971}
we split the pair potential \pot{}, as it is commonly done 
\cite{Hansen-et:1976}, into a purely attractive contribution
\begin{equation}
\label{eq:eff_attraction}
    \pot[a]\of{r}=
      \begin{cases}
       \pot\of{r_{min}},&r\leq r_{min} \\
       \pot\of{r},& r_{min} < r,
      \end{cases}
\end{equation}
where $\pot$ has its minimum at $r_{min}$, and into an effective  HS core,
$\exp\set{-\pot[HS]\of{r}}=H\of{r-\sigma}$, with the diameter $\sigma$ defined by
\begin{equation}
\label{eq:eff_hs_diameter}
   \sigma= \int_0^{r_0}\of{1-\exp\set{-\pot\of{r}}} \text{ d}r,
\end{equation}
where $\pot\of{r=r_0}=0$. 
We have checked that for the potential \pot{} considered here (c.f., 
\eref{eq:coll_potential}) and for its {\it s}plit  potential
\begin{equation}
\label{eq:pot_splitted}
    \pot[s]\of{r}=\pot[HS]\of{r}+ \pot[a]\of{r},
\end{equation}
as given by Eqs.~\eqref{eq:eff_attraction} and
\eqref{eq:eff_hs_diameter}, the resulting values of 
\bsec{} are almost the same.

\subsubsection{Density functional theory \label{ssc:coll_dft}}

Density functional theory (DFT) is based on the fact that there is a 
one-to-one correspondence between the local equilibrium number density 
$\rho\of{\x}$  of a fluid and a spatially varying external potential 
acting on it.
It follows that there exists a unique functional $\Fe\fd{\rho}$ which
is minimized by the equilibrium one-particle number density and that
the equilibrium free energy of the system is equal to the  minimal
value of the functional \cite{Evans:1979}. 
Since for static properties the absolute size of the particles does not 
matter, DFT has turned out to be very successful not only in 
describing simple fluids but also colloidal suspensions \cite{DFTcolloids}.
While the ideal gas contribution to the functional \Fe{} is known exactly,
$\Fe[id]\fd{\rho}/(k_BT) =
    \int \text{d}\x\; \rho\of{\x} \set{\ln\fd{\lambda^3\rho\of{\x}}-1}$,
where  $\lambda$ is  the  thermal wavelength,
the expression for the excess contribution $ \Fe[ex]=\Fe-\Fe[id]$
is known only approximately.

As a first step, for a liquid in a volume $v$ we consider
the simple functional
\begin{equation}
 \label{eq:dft_rpa}
 \begin{split}
    & \Fe[ex]\fd{\rho}= \\ 
 	& \Fe[HS]\of{\fd{\rho}; \sigma} +
	 \frac{k_B T}{2}\iint\limits_{v}
	 \text{d}^3\x[1]\text{d}^3\x[2]\;
	 \rho\of{\x[1]} \pot[a]\of{r_{12}}\rho\of{\x[2]},
 \end{split}
\end{equation}
where $r_{12}=\abs{\x[1]-\x[2]}$.
For the attractive contribution $\pot[a]$ to the interaction potential 
entering into \eref{eq:dft_rpa} we employ 
\eref{eq:eff_attraction}.
$\Fe[HS]\of{\fd{\rho}; \sigma}$ is the excess functional
for the HS system for which we consider the effective HS diameter 
$\sigma$ as given by \eref{eq:eff_hs_diameter}. 
For $\Fe[HS]$ various sophisticated functionals 
are available
\cite{Rosenfeld:1989,Roth-et:2002,HansenGoos-et}.
For our purposes, however, it is sufficient to determine  the 
{\it bulk} free energy density. 
According to \eref{eq:dft_rpa} the so-called random phase 
approximation for \Fe{} is given  by   
\begin{widetext}
\begin{equation}
\label{eq:febulk_pyrpa}
      \frac{\pi\sigma^3/6}{v}\Fe[RPA]/\of{ k_B T}=
       \eta_{\sigma} \of{\ln\of{\frac{\eta_{\sigma}}{1-\eta_{\sigma}}}
       -\frac{2-10\eta_{\sigma}+5\eta_{\sigma}^2}{2\of{1-\eta_{\sigma}}^2}}
       +\frac{1}{2}{{\eta_{\sigma}^2}}\widetilde{\pot}_{a,0}
\end{equation}
\end{widetext}
where for the HS-contribution in \eref{eq:febulk_pyrpa} 
the PY-approximation (as obtained via the compressibility 
equation) has been adopted \cite{Hansen-et:1976}; 
$\widetilde{\pot}_{a,0}=\frac{6}{\pi\sigma^3}\ft{\pot}_a(q=0)$ with 
$\ft{\pot}_a\of{q}$ as the Fourier transform of
the potential. The packing fraction $\eta_{\sigma}$ used in
\eref{eq:febulk_pyrpa} corresponds to the effective HS system, i.e.,
$\eta_{\sigma}=\frac{\pi}{6}\sigma^3\rho=\of{\frac{\sigma}{2R}}^3\eta$.
 In \eref{eq:febulk_pyrpa}
$\Fe[id]/\of{k_B T}=
    v\rho\of{ \ln\of{\eta_{\sigma}}
	     -\ln\of{\frac{\pi}{6}\of{\sigma/\lambda}^3}
	     -1}$
is incorporated, except for  the term
$v\rho \ln\of{\frac{\pi}{6}\of{\sigma/\lambda}^3}$ which accounts only
for a shift of \Fe{} linear in $\rho$ and therefore is irrelevant for 
determining phase coexistence. For the free energy given in 
\eref{eq:febulk_pyrpa} the critical point is given implicitly by
$\eta_{\sigma,c}^{\of{RPA}}=0.129$ and
$\of{\widetilde{\pot}_{a,0}}_c=-21.3$.
By choosing the PY-approximation we are able to compare the
results obtained by DFT and by the IEA on the same  footing.

\subsection{Critical phenomena in confined geometries} \label{ssc:critphen}

In this  subsection, within the field-theoretical framework we 
provide the relevant theoretical background for  critical phenomena 
in confined geometries. We pay special attention to the case of a 
binary liquid mixture close to its demixing point.

\subsubsection{General concept }\label{subsec:crit_generals}

The free energy \Fe{} of a system  (i.e., the {\it s}olvent in the case 
considered here, see below) 
close to its critical point \of{\tcb,c_{a,c}^{(s)}} is the sum
of a {\it r}egular, analytic background contribution \Fe[r] and a
{\it s}ingular part \Fe[s]. Within the approach of the field-theoretical
renormalization group theory the leading behavior
of the singular contribution \Fe[s] for a confined system is captured by the
(dimensionless) effective Landau-Ginzburg Hamiltonian
\begin{equation}
\begin{split}
  \label{eq:lgh}
  \Ham\fd{\Op}=& 
  \int_{v}\set{\frac{1}{2}\of{\nabla \Op}^2+\frac{\tau}{2}\Op^2
	+\frac{g}{4!}\Op^4-\hb\Op}\text{d}^{d}{ \x} \\
  &+\int_{\dv}\set{\frac{\c}{2}\Op^2-\hs\Op}\text{d}^{\of{d-1}}{ \x}, 
\end{split}
\end{equation}
where $v$ is the volume available to the critical medium 
and \dv{} is its confining \of{d-1}-dimensional surface.
$\Op=\Op\of{\x}$ is the local order parameter (OP) field associated 
with the phase transition. For the case of binary liquid mixtures, 
which in their bulk belong to the Ising UC, the OP
$\Op\sim \of{c_{a}-c_{a,c}^{(s)}}/c_{a,c}^{(s)}$ is a scalar.
The quartic term with the coupling constant $g>0$ stabilizes the 
Hamiltonian for $0>\tau\propto t$ ($t=\pm\of{T/\tcb-1}$, see 
the text below \eref{eq:coll_potential0}) and \hb{} is a
symmetry breaking {\it b}ulk field conjugate to \Op{}.
In \eref{eq:lgh} the surface couplings encompass the so-called surface 
enhancement  $\c{}\of{\x}$ and the  {\it s}urface field $\hs{}\of{\x}$ 
with $\x\in\dv$.
Equation (\ref{eq:lgh}) turns out to capture the fixed-point Hamiltonian 
of surface critical phenomena \cite{Diehl:1986}. For laterally inhomogeneous 
substrates \hs{} and \c{} vary along  \dv{} \cite{Troendle-et:2010}.
For colloids strongly preferring one of the two species of the binary 
liquid mixture the so-called strong adsorption limit applies, which is 
described by the so-called normal fixed  point,
i.e., ($\hs=\infty$, $\c=0$)${}^{}$ for all $\x\in\dv$.
Corrections to the   fixed-point  behavior of the CCF due to finite
surface fields and the crossover between various surface universality 
classes have been studied in detail for the film geometry in  $d=3$ 
by    MC  simulations
\cite{Hasenbusch-cross,vas-cross},
in $d=2$ by using exact solutions
\cite{Abraham-et:2010,Nowakowski-et:2009}, 
and in  $d=4$ within Landau-Ginzburg theory \cite{Mohry-et:2010}.
In a systematic perturbation theory in terms of $\epsilon=4-d$ (with 
$d^*=4$ as the upper critical dimension for the bulk Ising UC) thermal 
fluctuations are captured with a statistical weight  \cite{Diehl:1986} 
$\sim \exp \of{-\Ham\fd{\Op}}$.
MFT corresponds to the lowest order \of{d=4} in this 
expansion and accordingly the MFT-equilibrium
configuration $\Op[MFT]$ minimizes $\Ham\fd{\Op}$.

\subsubsection{The critical Casimir force}
\label{subsec:crit_ccf}

For the case of colloidal suspensions considered here, the colloids act 
as cavities  in the critical medium and thus in \eref{eq:lgh} the confining 
surface \dv{}  is the union of the surfaces of 
all colloids in the system. Accordingly, the OP profile for a 
given colloid configuration depends on the position of all colloids and therefore 
the CCFs which act on the confining surface \dv{}  are non-additive.
In order to cope with this very demanding challenge, here
we restrict our analysis to such low number densities $\rho$ 
of the colloids that the mean distance $\rho^{-1/d}$ between the 
colloids is large compared with the range of the CCFs, i.e., 
$\rho^{-1/d}> 2R+\xi$. In this limit the approximation
of pairwise additive CCFs is expected to be valid.

For the  pair potential $V_c^{\of{d}}=k_{B} T  \ccp^{\of{d}}$
of the CCF scaling theory predicts
\begin{equation}
  \label{eq:ccf_spsp_scaling_org}
 \ccp^{\of{d}}\of{D;R,t,\hb} 
		= \Delta^{-1}\widetilde{\theta^{(d)}}
		\of{\Theta,\Delta,\Lambda} ,
\end{equation}
where $\widetilde{ \theta^{(d)}}$ is a universal scaling function  
and the scaling variables are
$\Theta=\sgn\of{t} D/\xi$, $\Delta=D/R$, and
$\Lambda=\sgn\of{\hb}D/\xi^{\of{h}}$.
Within MFT for the Hamiltonian given in \eref{eq:lgh} one has
\cite{colloids1b} $\xi\of{t>0}=\abs{\tau}^{-1/2}$ and 
$\xi^{\of{h}} =3^{-1/2}\abs{\sqrt{g/6} \hb}^{-1/3}$.
$\ccp^{\of{d}}$ depends on the sign of $\Lambda$ because the surface 
fields $\hs$ imposed by the colloids break the bulk 
symmetry w.r.t. $\hb\to-\hb$.
For the following study it is useful to introduce another 
scaling variable ${\Sigma=\Lambda/\abs{\Theta}}$,
which depends solely on the properties of the solvent but not on the 
surface-to-surface distance $D$ between the colloids. Therefore, 
$\ccp^{\of{d}}$ is given by another, also universal scaling function
(compare \eref{eq:coll_potential0})
\begin{equation}
  \label{eq:ccf_spsp_scaling}
  \ccp^{\of{d}}\of{D;R,t,\hb} 
		= \Delta^{-1}
		\sfp[\of{d}]\of{\Theta,\Delta,\Sigma}, 
\end{equation}
with the scaling variable
$\Sigma=\sgn\of{\hb}\xi/\xi^{\of{h}}$.

In $d=3$ the scaling function \sfp[\of{d=3}] is  not known for the full 
range of the scaling variable $\Delta$ and  there are no results 
concerning its dependence on $\Sigma$.
However, if the colloid radius is large compared  to $\xi$ only those 
surface-to-surface distances $D$ matter which are small compared with $R$. 
This implies the validity of the Derjaguin approximation which allows us 
to express  \sfp[\of{d}] in terms of the universal
scaling function for the film geometry. The latter is known from MC 
simulations \cite{vas,Hasenbusch} in $d=3$ and for $\hb=0$.
In the opposite, so-called protein limit $D\gg R$ additional knowledge 
about the CCFs is available \cite{colloids1a,colloids1b}.
For the configuration of a colloid near a planar wall it has been  found
\cite{Hertlein-et2008,Gambassi-et:2009,Troendle-et:2010} that the 
Derjaguin approximation is valid up to $\Delta=D/R\lesssim 0.3$. 
Concerning a discussion of the Derjaguin approximation 
for the sphere-sphere geometry see Ref.~\onlinecite{colloids1a}
and in the case of non-spherical objects near a  
wall see Ref.~\onlinecite{khd-08}.
For the configuration of two spheres as studied here 
one has \cite{Derjaguin:1934,colloids1a,Gambassi-et:2009}
\begin{equation}
\label{eq:sfpot_derjaguin}
	\sfp[\of{d=3}]_{Derj}\of{\Delta,\Theta,\Sigma}= 
		\pi \int_1^{\infty}\of{x^{-2}-x^{-3}}
		\sff[\parallel]^{\of{d=3}}
		     \of{x\Theta,\Sigma} \text{ d}x,
\end{equation}
where  $\sff[\parallel]^{\of{d}}\of{y,\Sigma}$ 
is the universal scaling function of the CCF \ccf[,\parallel] per 
$k_B\tcb$ and per area $A$ for a slab of the thickness $L$:
$\ccf[,\parallel]/A=k_{B} \tcb L^{-d}
      \sff[\parallel]^{\of{d}}\of{y=\sgn\of{t}L/\xi,\Sigma}$.
We point out that within the Derjaguin approximation the scaling function \sfp[\of{d=3}] does
not depend on $\Delta$, which therefore enters into 
\eref{eq:ccf_spsp_scaling} only as a prefactor.

There  are experimental indications \cite{Beysens-et:1985,Nellen_hdependence} 
and theoretical evidence \cite{colloids1b} for a pronounced dependence of 
CCFs on the composition of the binary liquid  mixture acting 
as a solvent. As discussed in Appendix~\ref{app:1} this translates 
into  the dependence  on the scaling variable $\Sigma$.
In order to be able to capture this dependence to a certain  extent on
the basis of presently available theoretical knowledge  we propose the 
following approximation for $\sff[\parallel]^{\of{d}}\of{y,\Sigma}$:
\begin{equation}
\label{eq:sfp_h_approx}
	\sff[\parallel]^{\of{d}}\of{y,\Sigma}\simeq
		\sff[\parallel]^{\of{d}}\of{y,\Sigma=0}
		\frac {\sff[\parallel]^{\of{d=4}}\of{y,\Sigma}}
		      {\sff[\parallel]^{\of{d=4}}\of{y,\Sigma=0}}.
\end{equation}
This approximation offers three advantages:
(i) For $d\to 4$, i.e., for MFT, the rhs of \eref{eq:sfp_h_approx} reduces 
to the correct expression for the full  ranges of all scaling variables.
(ii) For $\hb\to 0$ the rhs of \eref{eq:sfp_h_approx} reduces 
exactly to $\sff[\parallel]^{\of{d}}\of{y,\Sigma=0}$ for all $d$.
In a certain sense the MFT approximation is
concentrated in the dependence on \hb.
(iii) The MFT treatment of the dependence on \hb{} does not suffer from
not knowing the amplitude of $\sff[\parallel]^{\of{d=4}}$
within this approximation;
for $d<4$  on the rhs of \eref{eq:sfp_h_approx} this amplitude drops out.
Within the MFT expressions in  \eref{eq:sfp_h_approx} the scaling variables 
$\Theta$  (entering via \eref{eq:sfpot_derjaguin}) and $\Sigma$ are taken 
to involve the critical bulk exponents in spatial dimension $d$ so that
the approximation concerns only the shape of the scaling function itself 
which typically depends on $d$ only mildly (see, e.g., the comparison 
of MFT results with corresponding results obtained by MC 
simulations for $d=3$ or with exact results for $d=2$ in
Refs.~\onlinecite{vas,Mohry-et:2010}).
Here we have calculated $\sff[\parallel]^{\of{d=4}}$ also for $\Sigma\neq0$.
For $\sff[\parallel]^{\of{d=3}}\of{y,\Sigma=0}$ we use the
MC simulation results \cite{vas}.
The scaling functions resulting from \eref{eq:sfp_h_approx} and from 
the local functional method \cite{upton,FdeG_loc_fun} are 
comparable \cite{tobepublished} and in qualitative agreement with
corresponding curves provided in Ref.~\onlinecite{Buzzaccaro-et:2010}. 

It is most suitable to determine the CCF from \Op[MFT] via the so-called 
stress-tensor \T{} \cite{Eisenriegler-et:1994}. The CCF per area in a 
slab, which is confined along the $z$-direction, is given by the \of{z,z} 
component of the thermally averaged stress tensor,
$\ccf[,\parallel]^{\of{MFT}}/A=
    {k_B \tcb}
    \thermal{\T[z,z]\fd{\Op[MFT]}-\T[z,z]\fd{\Op[b,MFT]}}$, with
\begin{equation}
\label{eq:stress_slab}
	\T[z,z]\fd{\Op} = 
	       \frac{1}{2}   \of{\Op'}^{2}
	      -\frac{\tau}{2}\Op^{2}
	      -\frac{g}{4!}\Op^{4}+\hb\Op,
\end{equation}
where the OP $\Op=\Op\of{z=z_0}$ and its derivative 
$\Op'=\of{\partial\Op/\partial z}_{z=z_0}$ 
are evaluated at an arbitrary point  $-L/2\leq z_0 \leq L/2$
within the slab and $\Op[b]$ is the bulk OP.
(The so-called  ``improvement`` term  of 
the canonical  stress-tensor \cite{Krech:1990:0} can be neglected 
because it does not contribute to the CCF.)

\section{Results \label{sec:results}}

\subsection{Range of parameters} \label{ssc:res_range}
In our study we use the following pair potential $\pot$
(see Eqs.~\eqref{eq:coll_potential0} and \eqref{eq:sfpot_derjaguin}):
\begin{widetext}
\begin{equation}
\label{eq:coll_potential}
	\pot\of{r=D+2R}=
		\begin{cases}
		\infty & x<0 \\
		s \set{a \exp{\of{-x}}
		  +(1/x) \sfp[\of{d=3}]_{Derj}\of{x/\zeta,\Sigma} }
		   & x>0,
		\end{cases}
\end{equation}
\end{widetext}
where $s=\kappa R$, $a=A/s$, $\zeta=\sgn\of{t}\kappa \xi$, and 
$x=\kappa D=\kappa r -2 s$. (The amplitude $a$ should not be confused 
with the acronym for the preferentially adsorbed  phase.) 
This parametrization has the advantage, that the shape of \pot{} is 
determined by $a$, $\zeta$, and $\Sigma$, while $s$ tunes the overall 
strength of the potential without affecting its shape.
The ratio of the competing length scales of repulsion and of the  CCF 
are measured by $\zeta$, which is typically varied experimentally;
$a$ is usually kept constant and provides a measure of the repulsion,
while $\Sigma$ (\eref{eq:ccf_spsp_scaling}) depends solely on the
thermodynamic state of the solvent.

In the following we discuss the ranges of the values
of the parameters entering into the effective potential
(\eref{eq:coll_potential}) and of the scaling variable $\Sigma$
which correspond to possible experimental realizations.
The radius $R$ of colloidal particles typically varies between
 $ 0.1\mu m $ and  $1\mu m $.
In the experiments reported in the present context so far colloidal
suspensions had been stabilized by electrostatic repulsion, with the
value of the strength parameter $A$ ranging over several orders of
magnitudes, i.e., from $A\simeq 10^2$ up to $10^{5}\ldots10^{6}$
(see, e.g., Refs.~\onlinecite{Hertlein-et2008,Gambassi-et:2009,Kurnaz}).
The value of  $\kappa^{-1}$ can be tuned  by salting the solution.
Due to screening effects, an increased  amount of ions in the solution
leads to a decrease of  $\kappa^{-1}$. Although the coupling between
the charge density and  OP fluctuations is not yet fully understood,
there  is experimental \cite{Nellen-et:2011} and theoretical
\cite{Ciach-et:2010,Bier-et:2010} evidence that  critical adsorption and
CCFs can be  altered significantly by adding ions to the  binary liquid mixture.
Such subtle mechanisms are not taken into account within the
effective potential discussed here (\eref{eq:coll_potential}).
Therefore it is applicable only for not too small values of the
screening length, i.e., for $\kappa^{-1}\gtrsim 10 nm$.
For  binary liquid mixtures the correlation length amplitude
$\xi_0$ is of the order of few {\AA}ngstrom.
The relevant experiments have been carried out at room temperature
due to $\tcb   \approx  300 K$.
In those experiments deviations from the critical temperature
as small as  $T-\tcb{}\sim 10 mK$ have been resolved 
\cite{Hertlein-et2008,Gambassi-et:2009},
which corresponds to a correlation length $\xi$ of a couple of tens of $nm$.

It is  more difficult to assess the experimentally relevant range of the
scaling variable $\Sigma$,  which  is a function of the bulk ordering field
\hb. Often the amplitude $\xi^{\of{h}}_0$ of the correlation length
$\xi^{\of{h}}\of{\hb}$ is not known. However, one may use the
equation of state (\eref{eq:eos_critical_scaling})
which relates  $\Sigma$ to the scaling variable $X$ associated with
the OP \Op{} (see the text after \eref{eq:x}).
In terms of the  parameters of the potential given in
\eref{eq:coll_potential},  for $t>0$ one has  
$X=m_0 \zeta^{-1/\nu}$ with (see Appendix~\ref{app:1})
\begin{equation}
\label{eq:op_amplitude_modified}
m_0=\sgn\of{\phi}\of{\zeta_0^{+}}^{1/\nu}\abs{\mathcal{B}/\phi}^{1/\beta}
\end{equation}
and $\zeta_0^{+}=\kappa \xi_0^{+}$.
For example, in binary liquid mixtures the order parameter \Op{} is
proportional to the deviation  of the concentration
$c_a$ of the component $a$ from its critical value $c_{a,c}^{(s)}$,
$\Op= {\cal A}(c_a-c_{a,c}^{(s)})$ (note that $\mathcal B$ is
proportional to ${\cal A}$),  which can be easily controlled by changing
the mass or the volume fraction of one of the components of the mixture.
The experiments reported in Refs.~\onlinecite{Hertlein-et2008,Gambassi-et:2009}
provide indications  concerning the size of the critical region in the 
thermodynamic direction orthogonal to the temperature axis for the
binary liquid mixture of water and lutidine near the consolute point 
of its phase segregation.
These measurements revealed the occurrence of CCFs within the range 
of the lutidine mass fraction $\m[L]$ deviating from its critical 
value $\m[L,c]$ up to $\abs{\m[L]-\m[L,c]} \simeq 0.04$.
From the experimental data in
Refs.~\onlinecite{Handschy-et:1980,Jacobs-et:1977,Jayalakshmi-et:1994} 
one finds \cite{WL-note} for the water-lutidine mixture
${\mathcal{B}_{\m{}}}\approx  1.0$. The 
index $\m$ refers to the specific choice of the order parameter, i.e., 
$\Op[\m]\of{t\to0^{-}}\equiv \m[L]-\m[L,c]
\equiv\mathcal{B}_{\m}\abs{t}^{\beta}$.  
Fitting the experimentally determined coexistence curve 
\cite{Beysens-et:1985,Kurnaz} yields a somewhat smaller value 
$\mathcal{B}_{\m}=0.765$ which we adopt in the following.
Therefore the difference  $\abs{\m[L]-\m[L,c]} \simeq 0.04$ 
corresponds to $\abs{\mathcal{B}/\Op}=19$, which for, e.g.,
$\zeta_0^+ = 0.02$  translates into  $\abs{m_0}\simeq 17$
(\eref{eq:op_amplitude_modified}), where we have used the critical
exponents of the three-dimensional Ising universality class \cite{Pelissetto-et:2002}: 
$\nu\of{d=3}=0.6301(4)$ and $\beta\of{d=3}=0.3256(3)$. 
(The critical composition corresponds 
to $m_0=\pm \infty$.) Accordingly, for the
temperature differences accessible in these experiments, i.e., for
$\of{T-\tcb{}}/\tcb{} \approx 3\times10^{-5}$,
one has for the scaling variable $\abs{X}\approx 0.26$,
corresponding to $\abs{\Sigma}\approx 6.6$.

As discussed  before, the effective pair potential given in
\eref{eq:coll_potential} is applicable only for  sufficiently large
distances $D\gtrsim\kappa^{-1}$ because it takes only the interactions
of the double-layers into account and neglects possible short-ranged
contributions to effective  van-der-Waals interactions.
Furthermore, the critical Casimir potential takes its universal form
(\eref{eq:ccf_spsp_scaling}) only in the scaling limit, i.e., for distances $D$ 
which are sufficiently large compared with the correlation length amplitude 
$\xi_{0}^{+}\approx 0.25nm$.  Analogously also $\xi$ and $\xi^{\of{h}}$
must be sufficiently large compared with microscopic scales.
Later on, in order to circumvent the unphysical divergence
$\sim x^{-1}$ in $\pot\of{D\to 0}$ (see \eref{eq:coll_potential}) for small
distances $D$ we shall consider, as far as necessary, also  a linear extrapolation:
\begin{equation}
\label{eq:lin_extrapolation}
    \pot{}\of{D<D_0}= 
	A\exp{\of{-\kappa D}}+ \pot[c,0]+\of{D-D_0} \pot[c,0]',
\end{equation}
where $D_0\simeq \xi_0$, $\pot[c,0]= \pot[c]\of{D=D_0}$,
and $\pot[c,0]'= \of{\partial \pot[c] /\partial D}_{D=D_0}$.

\subsection{Thermodynamics \label{ssc:res_thermo}}
\begin{figure*}
	\includegraphics{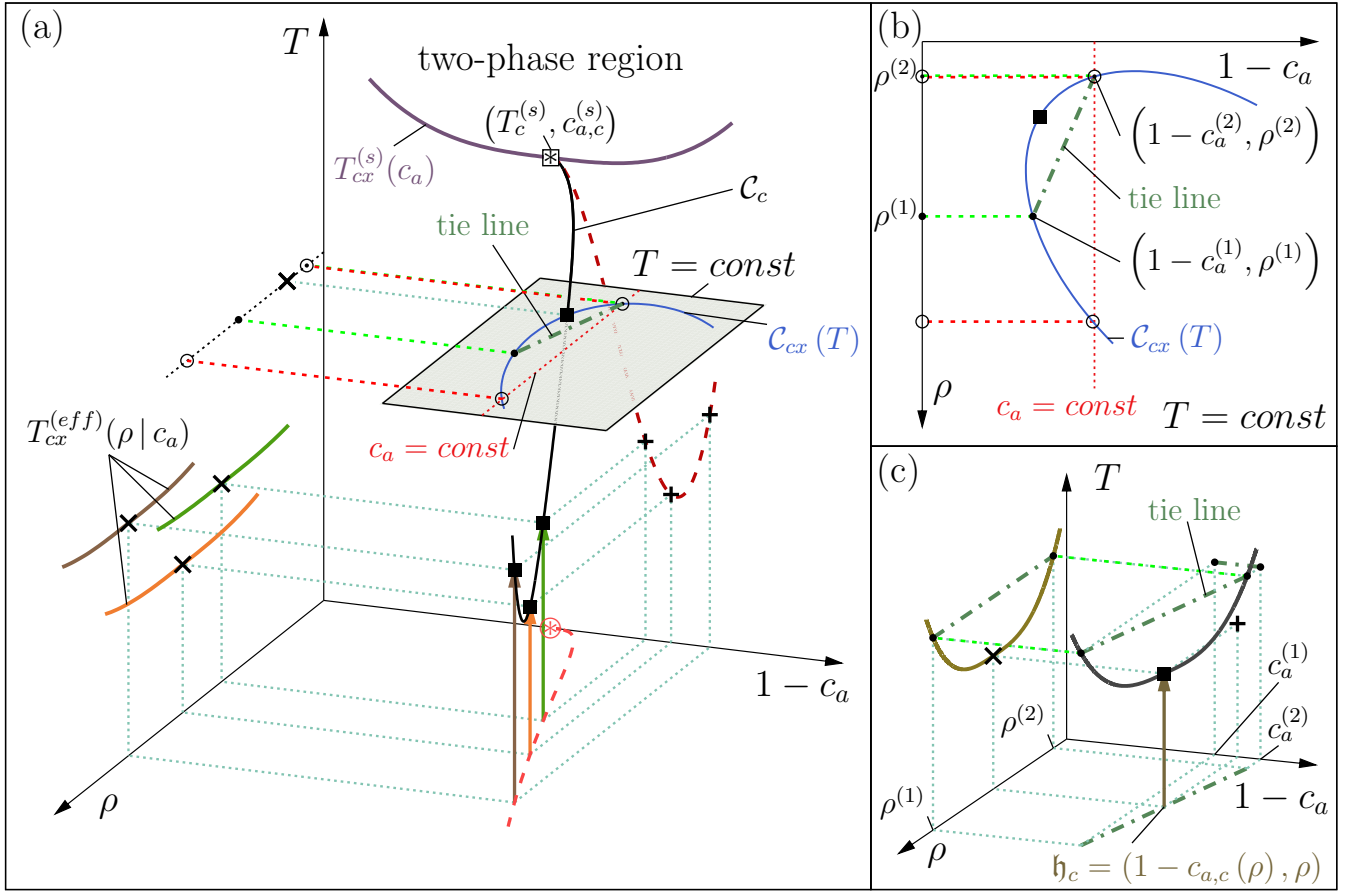}
	\caption{
	Sketch of the phase diagram for colloids immersed in a binary   
	liquid mixture, the latter exhibiting a closed-loop
        miscibility gap. We focus on the region around the 
	lower critical point $\boxast$ ($\tcb,c_{a,c}^{(s)},\rho=0$)
	(see the lavender phase separation curve 
        $T^{\of{s}}_{cx}\!\of{c_a}$ in the \of{T,c_a,\rho=0} plane).
	For fixed pressure corresponding to a liquid state of the
        system, the thermodynamic (td) space of the pure solvent 
        is two-dimensional; here we consider as variables 
        the temperature $T$ and the concentration $c_a$. Upon adding 
        colloids at fixed pressure the td space becomes 
	three-dimensional; we use the colloidal number density $\rho$ as 
	additional td variable. The miscibility gap of the pure solvent 
	in the $\of{T,c_a,\rho=0}$-plane extends to a two-phase region 
	in the three-dimensional td space and is bounded by a
	two-dimensional manifold $T_{cx}\of{c_a,\rho}$ of 
	{\it{}c}oe{\it{}x}isting states (not shown).
	The tube-like shape of $T_{cx}\of{c_a,\rho}$ is not straight 
	but bent and twisted. Its actual form is expected to 
	depend sensitively on all interactions (i.e., 
        the solvent-solvent, the solvent-colloid, and the 
        colloid-colloid interactions). Each state on 
        $T_{cx}\of{c_a,\rho}$ coexists with another one, both being 
        connected by a horizontal and straight, so-called 
        tie line (green dash-dotted line).
	There is a line \cur[c]{} (black curve) of critical points 
	(some of which are shown as black squares) embedded in 
	$T_{cx}\of{c_a,\rho}$ which is the extension of the critical 
        point \of{\tcb,c_{a,c}^{(s)},\rho=0}. \cur[c]{} is given by the
        states for which the tie line has zero length. 
	The projections of \cur[c]{} (with selected critical 
        points \Rectsteel{}) onto the planes \of{\rho,c_a} and \of{T,c_a} 
        (with $\boldsymbol{\mathrm +}$ as the projection of \Rectsteel{}) are 
	indicated as red dashed curves. 
	For $\rho\to 0$ the bending of the curve \cur[c]{} can be 
        inferred from scaling arguments (\eref{eq:cp_shift}). 
	For colloidal suspensions interacting via an
        effective potential which consists of a 
	soft repulsion and the critical Casimir attraction, 
	the results of the effective approach suggest that,
	for intermediate values of $\rho$, \cur[c]{} bends up again 
	(see Subsect.~\ref{ssc:pd}). This bending is due to the specific 
	properties of the critical Casimir forces and is indicated in (a). 
	Effective colloidal models render coexistence curves
	$T^{{(eff)}}_{cx}\!\of{\rho \,\vert\, c_a}$ which are explicit
	functions of $\rho$ only and depend parametrically on the overall
	concentration $c_a$ (see (a) for three examples; the three
	vertical arrows indicate thermodynamic paths describing the
	approach of the corresponding critical point \Rectsteel{} 
        upon raising temperature).
	For suitable effective models and within a certain 
        region of the td space (see main text), the critical points 
        ($\boldsymbol{\times}$), \of{T^{{(eff)}}_{c}\of{c_a},\rho_c\of{c_a}},
	of the coexistence curves 
        $T^{{(eff)}}_{cx}\!\of{\rho \,\vert\, c_a}$ are expected to 
        approximate the projection of \cur[c]{} onto the \of{T,\rho}-plane. 
	Within the effective approach a unique value $c_a$ is 
	taken throughout the whole system. 
	In contrast, all three panels show that in general for $T=const$ 
        the coexisting phases (i.e., the points connected by a tie-line) 
        differ both in $\rho$ {\it{}and} $c_a$. Thus the 
	effective approach has a limited applicability for determining 
	the phase diagram. Experimentally, upon increasing 
        temperature one is able to determine a (in general 
        nonplanar) coexistence curve (black line in (c)) in 
	the three-dimensional td space. In (c), for a selected 
        critical value $\hSt[c]=\of{1-c_{a,c}\of{\rho},\rho}$ 
        (i.e., \hSt[c] is a point on the red dashed line
	in the \of{\rho,c_a}-plane) such a curve (black line), 
        its projection onto the \of{T,\rho}-plane (mustard line), 
        as well as a selected tie line and its three projections 
        are shown. Note that the tie line is not parallel to the 
        $\rho$-axis. This means that the coexisting phases differ 
        with respect to $c_a$, in contrast to the aforementioned 
        assumption of an unique value of $c_a$ for the effective 
        one-component description. Thus the mustard curve in (c) will 
        in general differ from the corresponding curve
        $T^{{(eff)}}_{cx}\!\of{\rho \,\vert\, c_a}$ in (a), even if the 
        associated critical points are the same.
    } 
    \label{fig:phases_sketch}
\end{figure*}
In this section, we consider pair potentials which are 
repulsive at short distances (i.e., $a$ is sufficiently 
large) so that the suspension is stable.

\subsubsection{General discussion}\label{subsec:td_generals}

In this Subsection we consider the thermodynamics of 
actually ternary colloidal suspensions with binary solvents such as
water-lutidine mixtures which exhibit a closed-loop two-phase 
region of demixed phases (each being rich in one of the two species). 
We focus on that region of this miscibility gap 
which is close to the lower critical point. 
For fixed pressure, their thermodynamic states can be characterized
by the temperature $T$ and the concentration $c_a$ of 
one of the species with the critical  point \of{\tcb,c_{a,c}^{(s)}} and the
liquid-liquid phase {\it{}c}oe{\it{}x}istence curve 
$T^{\of{s}}_{cx}\!\of{c_a}$ in the absence of colloids. 
Upon adding colloidal particles to such a solvent, for fixed 
pressure the thermodynamic space of the system
becomes three-dimensional spanned by $T$, $c_a$, and, e.g., 
by the colloidal number density $\rho$ (see \fref{fig:phases_sketch}; 
one can also choose, instead, the fugacity of
the colloids).  Accordingly, the closed-loop 
phase coexistence curve $T^{\of{s}}_{cx}\!\of{c_a}$ becomes a 
two-dimensional, tubelike manifold $T_{cx}\of{c_a,\rho}$
with $T_{cx}\of{c_a,\rho=0}=T^{\of{s}}_{cx}\!\of{c_a}$. 
It contains a line \cur[c]{} of critical points
\of{T_c\of{\rho},c_{a,c}\of{\rho}} which is the extension of the
critical point of the solvent in the absence of colloids, 
i.e., $\of{T_c\of{\rho=0},c_{a,c}\of{\rho=0}}=\of{\tcb,c_{a,c}^{(s)}}$.
For  fixed temperature $T=const$, the set of pairs 
$\fd{\of{c_a^{\of{1}},\rho^{\of{1}}},\of{c_a^{\of{2}},\rho^{\of{2}}}}$ 
of coexisting states given by $T=T_{cx}\!\of{c_a,\rho}$ 
forms a curve $\cur[cx]\of{T}$ in the \of{c_a,\rho} plane,
the shape of which depends on the considered value of $T$
(see Figs.~\ref{fig:phases_sketch}(a) and (b)). In the phase 
diagram, two coexisting states are
connected by a straight, so-called tie line
(see \fref{fig:phases_sketch}). 
(If one chooses the fugacity of the colloids instead of $\rho$, 
one also obtains a tubelike manifold $T_{cx}$ of phase coexistence. 
However, in this case the horizontal tie lines lie in the plane of 
constant fugacity, i.e., parallel to the \of{T,c_a}-plane, because 
the coexisting phases share a common fugacity of the colloids.)

As stated above, the two-phase loop of the pure solvent 
(i.e., for $\rho=0$, bounded by $T^{\of{s}}_{cx}\!\of{c_a}$) extends 
into the three-dimensional thermodynamic space of the actual colloidal
suspension ($\rho \neq 0$). Due to the presence 
of additional interactions and degrees of freedom one expects that 
the shape of this two-phase region (bounded by 
$T_{cx}\!\of{c_a,\rho}$) is not a straight but a 
distorted tube. The actual shape 
of $T_{cx}\of{c_a,\rho}$ is expected to depend sensitively on all
interactions present in the ternary mixture, i.e., 
among the colloidal particles, between the colloidal and the
solvent particles, and among the solvent particles.
The relevance of the solvent-solvent interaction for the effective 
potential and, accordingly, for the phase behavior of the effective 
colloidal system has been demonstrated recently by 
MC studies in which various kinds of model solvents have  
been used \cite{Gnan-et:2011}. It is reasonable to expect that 
this relevance transfers also to the phase behavior of the full 
multi-component system. Such distortions of the phase diagram 
relative to that of the underlying binary mixture do occur for 
ternary mixtures of molecular fluids. For example, in 
Ref.~\onlinecite{Andon-et:1952} experimental studies of molecular ternary 
mixtures containing various kinds of lutidines are reported. 
These studies show that the upper and lower critical temperature 
for a closed-loop phase diagram can be tuned by varying 
only the concentration of the third component and that the two-phase 
loop can even disappear upon adding a third component. Similar 
experimental results are reported in Ref.~\onlinecite{Prafulla-et:1992}.  
Such complex phase diagrams can also be generated by adding 
colloidal particles to the binary solvent, as can be inferred from 
corresponding experimental studies 
\cite{Kline-et:1994,Jayalakshmi-et:1997,Koehler-et:1997}. 
In contrast to the  molecular ternary mixtures, for the latter kind of 
ternary mixtures a decrease of the lower critical temperature upon 
adding colloids as a third component is reported.

Theoretical studies of bona fide ternary mixtures have so 
far been concerned with, e.g., lattice gas models 
\cite{terMixTheory} and (additive or non-additive) mixtures of 
hard spheres, needles, and polymers \cite{Schmidt-et:2002,Schmidt:2011}. 
In these studies the constituents are of comparable size, i.e., 
their size ratios are less than ten.
The peculiarity of  the kind of mixtures considered here lies in the 
fact that the sizes of their constituents differ by a few 
orders of magnitude. This property distinguishes them significantly 
from mixtures of molecular fluids. In contrast to molecular ternary 
mixtures, in colloidal suspensions the colloidal particles influence 
the other two components not only by direct interactions but also 
via strong entropic effects. This is the case because  their surfaces 
act as confinements to fluctuations of the concentration of the 
solvent and they also generate an excluded volume for the solvent 
particles.  The importance of considering the colloidal suspension 
as a truely ternary mixture has been already pointed out 
in Ref.~\onlinecite{Sluckin:1990}.

\subsubsection{Scaling of the critical point shift} \label{ssc:sc}

For dilute  suspensions, i.e., for $\rho\to 0$, the shape of the line 
\cur[c]{} of the critical points can be estimated  by resorting to
phenomenological scaling arguments similar to the ones given by Fisher 
and Nakanishi \cite{Fisher-et:1981} for a critical medium confined between 
two parallel plates separated by a distance $L$. For a dilute 
suspension, the mean distance $\rho^{-1/d}$ between colloidal particles 
plays a role analogous to $L$. Close to the critical point of the 
solvent, due to $\xi\sim\abs{t}^{-\nu}$ one can identify the two relevant 
scaling variables $w^{\of{1}}\propto\abs{t}^{-\Delta}\hb$ and 
$w^{\of{2}}\propto \abs{t}^{\nu }\rho^{-1/d}$ (for the simplicity of the 
argument \cite{note_CP_scaling} here we do not consider the influence of 
the scaling variable $\abs{t}^{\nu}R$) and propose the scaling property
$\fe\of{T,\Delta\mu,\rho}\simeq \abs{t}^{2-\alpha}K\of{w^{\of{1}},w^{\of{2}}}$
of the free energy density (where  the difference $\Delta\mu\sim\hb$ of 
the chemical potentials  of the two components of the solvent acts as 
a symmetry breaking bulk field). The critical points are given by 
singularities in $K$ occurring at certain points 
\of{w^{\of{1}}_c,w^{\of{2}}_c}.  This implies that the critical point 
\of{T_c\of{\rho},\hb[,c]\of{\rho}} shifts according to 
\begin{subequations} 
  \label{eq:cp_shift}
  \begin{equation}
  \label{eq:cp_shift_hc-tc}
      T_c\of{\rho}-\tcb{} \sim \rho^{1/\of{\nu d}}
      \qquad  \text{and}  \qquad
      \hb[,c]\of{\rho}    \sim \rho^{\Delta/\of{\nu d}}
  \end{equation}
so that
  \begin{equation}
  \label{eq:cp_shift_cac}
      c_{a,c}\of{\rho}-c_{a,c}^{(s)} \sim \rho^{\Delta/\of{d\delta\nu}}
  \end{equation}
\end{subequations}
with $\of{1/\of{\nu d}}_{d=3}\simeq0.53$ and 
$\of{\Delta/\of{d\delta\nu}}_{d=3}\simeq 0.17$.
This states that in the presence of colloids the critical point 
occurs when the  bulk correlation lengths $\xi$ and
$\xi^{\of{h}}$ of the solvent become comparable 
with the mean distance $\rho^{-1/d}$  between the colloids.

\subsubsection{Effective one-component approach} \label{ssc:eff}
Integrating out  the degrees of freedom associated with the smallest 
components of the solution (here two) provides a manageable effective 
description of colloidal suspensions. This kind of approach is commonly 
used, for instance, recently for studying large particles 
immersed in various kinds of model solvents \cite{Gnan-et:2011} or 
in order to describe a binary mixture of colloids immersed in 
a phase separating solvent \cite{Zvyagolskaya-et:2011}.
However, this effective description has  only a limited range of 
applicability for investigating the phase behavior of colloidal 
suspensions. For example, it fails in cases in which the influence 
of a set of colloids on the solvent cannot be neglected or the 
(pure) solvent undergoes a phase seperation on its own, as considered 
here. In the effective approach, the concentration $c_a$ of one 
of the solvent species, averaged over the whole sample, enters as a 
unique  parameter into the effective interaction potential between the 
large particles (see the dependence on $\Sigma$ in
Eqs.~\eqref{eq:ccf_spsp_scaling} and \eqref{eq:coll_potential}). 
However, generically (e.g., due to the adsorption 
preferences of the colloids) the concentrations $c_a^{\of{i}}$ 
are different in the two coexisting phases $i=1,2$ 
(see \fref{fig:phases_sketch}) and thus, for a proper 
description, one would have to allow this parameter, and hence
the effective potential, to vary in space.
Experiments  have revealed \cite{gallagher:92,grull:97,note_col_in_solvent} 
that for suspensions very dilute in colloids 
with a phase separated solvent, basically all colloidal particles are 
populating the phase rich in the component preferred by the colloids
(with concentration $c_a^{\of{1}}$).
In this  case the  actual effective attractive interaction among the 
particles will be weaker than implied by the effective potential in 
which  only the {\it overall} concentration $c_a<c_a^{\of{1}}$ enters 
as a parameter. (We recall that the CCFs depend non-monotonously on 
$c_a$ and are strongest for $c_a\lesssim c_{a,c}^{(s)}$. Thus for
$c_a^{\of{2}}<c_a<c_a^{\of{1}}$ the CCFs differ for each
concentration and are - for the typical situation considered here - 
in general weakest for $c_a^{\of{1}}$; $c_a^{\of{2}}$ is the
concentration in the other coexisting phase.)
Within the effective approach for the colloids, one obtains, e.g., by 
means of DFT, coexistence curves 
$T^{{(eff)}}_{cx}\!\of{\rho \,\vert\, c_a}$ which depend parametrically 
on the  solvent composition $c_a \approx c_{a,c}^{(s)}$ (or equivalently, by
making use of \eref{eq:eos_critical_scaling}, on $\Sigma$ as in 
\eref{eq:coll_potential}). Since the effective potential corresponding
to this specific value of $c_a$ is used
both for the one-phase region as well as  for
both phases within the two-phase region, this implies that the  
tacitly assumed corresponding physical situation is such that
the composition $c_a$ is fixed  throughout the system
as an {\it external constraint}. In particular the two 
coexisting phases do not differ in their values of 
the concentration of the solvent particles but only in their colloidal
densities $\rho^{\of{1}}$ and  $\rho^{\of{2}}$. 
Thus within the effective approach for determining the phase 
behavior one of the essential features, i.e., the tendency of the 
solvent to phase separate, is suppressed.  Accordingly, it 
is not possible to construct the full coexistence manifold
$T_{cx}\!\of{c_a,\rho}$ on the basis of the curves
$T^{{(eff)}}_{cx}\!\of{\rho \,\vert\, c_a}$ alone. 
Rather, the effective approach is adequate  as long as to a large
extent the phase segregation involves only the colloidal
degree of freedom, i.e., the values of $\rho$ differ in the two phases,
but the values of $c_a$ are nearly the same. Therefore the approximation 
$T_{cx}\of{c_a,\rho}\approx T^{{(eff)}}_{cx}\!\of{\rho \,\vert\, c_a}$ 
is valid in a region of the thermodynamic space in which the actual 
tie lines happen to be almost orthogonal to the $c_a$-axis 
(see \fref{fig:phases_sketch}). 
For the phase diagram in terms of the variables $T$, $c_a$, and the 
fugacity of the colloids this latter condition implies that the tie
lines have to be sufficiently short.

We expect the  effective one-component approach to work well 
for temperatures corresponding to the one-phase region of the pure solvent 
and  for 
an intermediate range of values of  $\rho$. On one  hand $\rho$ should 
be {\it large enough} so that the competition between 
the configurational entropy and the potential energy due to the effective
forces  can drive a phase separation. On the other hand $\rho$ has to 
be {\it small enough} so that the approximation of 
using  an effective pair potential between the colloids is valid and the 
influence of the colloids on the  phase behavior of the solvent is 
subdominant.
Given such values of $\rho$ and $T$,  the reduced second
virial coefficient $\bsec^*$ (\eref{eq:stickyparam}) is an appropriate
measure for the strength of the attraction and a useful indicator of the
occurrence of a phase separation into a colloidal-rich (``liquid'') and a
colloidal-poor (``gas'') phase. According to the discussion above, for
values of the thermodynamic variables  (i.e., the values of $T$ and $\rho$,
and for the prescribed value of $c_a$) which are approximately 
the same as in the one-component approach, 
the ternary mixture is expected to exhibit a phase separation.
According to the VLNF conjecture the corresponding critical point 
should occur when $\bsec^*$ reaches the critical value
$\bsec[,c]^*\simeq-1.212$ (\eref{eq:stickyparam}) of Baxter's model.
While the VLNF conjecture provides an  empirical
estimate for the values of parameters for which there is a critical
point, within DFT  these parameters as well as the shape of the 
phase coexistence curve can be calculated. Within the effective 
approach only its dependence on the colloidal density number $\rho$ 
for  globally fixed values of $c_a$ can be determined.

\subsubsection{Phase diagrams}
\label{ssc:pd}

\begin{figure*}
	\includegraphics{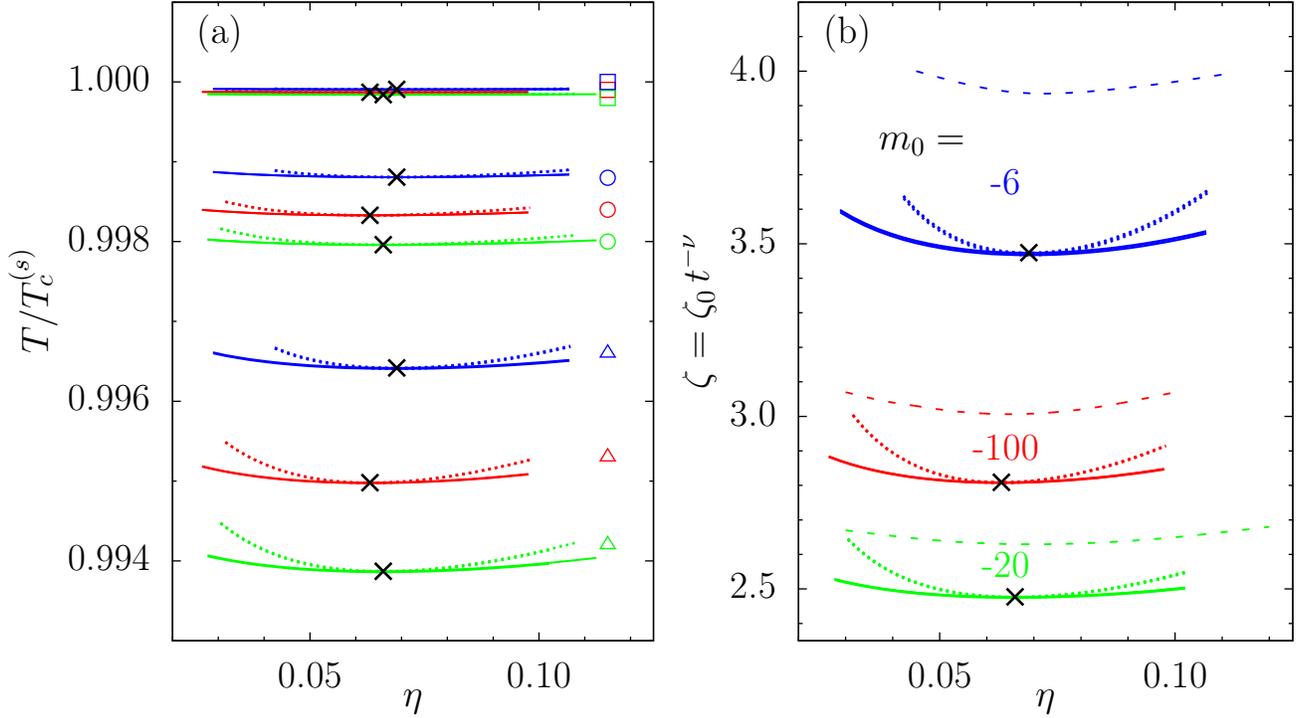}
	\caption{ 
	(a) Phase coexistence curves
	$T^{{(eff)}}_{cx}(\rho\,\vert\,c_a)$ (full lines), spinodals 
	(mean-field divergence of $\chi_T$, dotted lines), and the
	critical points $T^{(eff)}_{c}$ (crosses) of 
	an effective, one-component system of large particles
	(\eref{eq:coll_potential}) as obtained by DFT according 
	to \eref{eq:dft_rpa}. The curves correspond to a solvent with 
        a lower critical temperature \tcb{} and to various fixed solvent 
        compositions $c_{a}$ ($m_0= -100$ (red), $-20$ (green), 
        $-6$ (blue)). The other parameters are taken as  
	$\zeta_0=0.01\;\of{\Box{}}$, $0.05\;\of{\circ}$, and 
        $0.1\;\of{\vartriangle}$, $s=10$, and $aT/\tcb=100$.
	Close to the phase separation, the dominant temperature 
	dependence is that of the critical Casimir forces 
	(CCFs)  encoded in
	$\zeta\of{t=1-T/\tcb}=\sgn\of{t}\kappa\xi\of{t}
	    =\sgn\of{t}\zeta_0\abs{t}^{-\nu}$.
	Therefore, in terms of $\zeta$ (see
        panel (b)) the curves in (a) for different values of $\zeta_0$ 
        fall de facto on top of each other, independent of 
	whether the solvent has an upper or a lower critical point. 
	Within the integral equation approach only the spinodals have 
        been determined (PY: dashed lines in (b), see the main text). 
	According to \eref{eq:op_amplitude_modified},
	$\of{c_a-c_{a,c}^{(s)}}^{-1}\sim m_0= \pm \infty$ corresponds to the
	critical composition $c_{a,c}^{(s)}$. For solvent compositions  which
	are somewhat poor in the component preferred by the colloids, 
	i.e., for intermediate negative values of $m_0 \simeq -20 $, 
	the CCFs are strongly attractive. Therefore, for them small
	correlation lengths suffice to bring about phase separation;
	accordingly the binodals occur at small values of 
	$\zeta\sim\abs{t}^{-\nu}$. Here we consider only
	thermodynamic states of the solvent which 
	are in the one-phase region, i.e., $t>0$.
      }
      \label{fig:spinodals}
\end{figure*}

Generically, in experiments the solvent composition and hence the 
associated quantity $m_0$ (\eref{eq:op_amplitude_modified}) is 
fixed and $\zeta\sim\abs{t}^{-\nu}$ is varied. 
In Fig.~\ref{fig:spinodals}(a) for a solvent exhibiting a lower 
critical point and for the parameter choices 
$aT/\tcb=100$, $s=10$, $m_0=-100$, $-20$, and $-6$,
and $\zeta_0=0.01$, $0.05$, and $0.1$ we present the coexistence 
curves $T^{{(eff)}}_{cx}(\rho\,\vert\,c_a)$ 
and the spinodals of the colloidal ``liquid''-``gas'' phase transition 
as function of the colloid packing fraction $\eta=(4\pi/3)R^3\rho$ 
as obtained by the DFT presented in \eref{eq:dft_rpa}.  
(Note that in accordance with Eqs.~\eqref{eq:A_electric} and 
\eqref{eq:coll_potential} in the product $aT$ 
the explicit dependence on $T$ drops out.) 
For the corresponding  free energy \Fe[RPA] given in
\eref{eq:febulk_pyrpa} the effective colloidal system phase separates if 
$\widetilde{\pot}_{a,0}<\of{\widetilde{\pot}_{a,0}}_c=-21.3$. 
This condition is satisfied provided the attractive part of the 
interaction potential is sufficiently strong. We recall, that for the 
effective potential considered here attraction occurs for 
$T/\tcb=1\pm\of{\zeta/\zeta_0}^{-1/\nu}\to 1$ (see 
\eref{eq:coll_potential}). For a solvent exhibiting a lower 
(upper) critical point the lower (upper) sign holds 
in the one-phase region of the solvent, within which the effective 
approach is applicable. In this temperature limit
the variation of $a$ (Eqs.~\eqref{eq:A_electric} and
\eqref{eq:coll_potential}) with $T$ is subdominant and thus
the dependence of \Fe[RPA] on $T/\tcb$ and 
$\zeta_0$ reduces to a dependence on $\zeta$ only. 
Accordingly, as can be inferred from the comparison of 
Figs.~\ref{fig:spinodals}(a) and \ref{fig:spinodals}(b), in terms of
$\zeta$ for each value of $m_0$ the coexistence curves 
$T^{{(eff)}}_{cx}(\rho\,\vert\,c_a)$ for different values of 
$\zeta_0$ fall de facto on top of each other. 
The difference between these curves is of the order of 
$\of{\zeta_0/\zeta}^{1/\nu}$ (because 
$T/\tcb=1\pm\of{\zeta/\zeta_0}^{-1/\nu}$), which for the three values of 
$\zeta_0$ used in Fig.~\ref{fig:spinodals} is about the thickness 
of the lines shown in Fig.~\ref{fig:spinodals}(b). 
In terms of this presentation it does not matter 
whether the solvent exhibits a lower or an upper critical point. 
Although spinodals are mean-field artifacts, we present them nonetheless
because they provide some indication about the location of the binodal 
which encloses the former. Furthermore the spinodals carry the advantage 
that the isothermal compressibility $\chi_T$ is the property of only
one thermodynamic state. Therefore, in contrast to the calculation of 
the binodal (which depends on two coexisting phases), the calculation 
of the spinodal does not suffer from  the non-uniqueness of the 
effective potential in the case of phase coexistence. 
As discussed in Subsect.~\ref{ssc:coll_stability}, based on formally 
exact relations the phase behavior can in principle 
be calculated from the correlation functions obtained within the 
integral equation approach  (IEA).  
Within the so-called compressibility route (\eref{eq:compressibility})
the spinodals, i.e., the loci 
of the mean-field divergence of $\chi_T$, are directly accessible.
On the other hand, the binodals,  i.e., the 
loci of two thermodynamic states which at the same temperature 
have different packing fractions but the same pressure, are directly
accessible via the so-called virial route (\eref{eq:pressure_oz}). 
We refrain from calculating the binodal (the spinodal) 
via the compressibility route (virial route), because it would require 
thermodynamic integration, which we want to avoid due to the subtlenesses 
described in Subsect.~\ref{ssc:coll_stability}.
The spinodals calculated by the compressibility route are shown as 
dashed curves in \fref{fig:spinodals}(b). For the region of the 
thermodynamic space where the IEA renders solutions for the 
correlation functions, we failed to find binodals, i.e., along 
the pressure isotherms $p\of{\eta;T=const}$ as calculated via the virial 
route there are no two states with 
$\eta_1\neq \eta_2$ and $p\of{\eta_1;T}=p\of{\eta_2;T}$.
These observations can be explained by the thermodynamic inconsistency
of this  IEA. Due to the approximate bridge function, the binodals as 
obtained by two different routes need not to coincide.
The same observations are found within the HNC 
approximation within which the spinodals are shifted  w.r.t. to the PY 
results to slightly smaller values of $\zeta$. Although within the IEA 
the binodals could not be determined, at least the loci 
of the obtained spinodals are similar to the ones obtained from DFT.

Since the CCFs  are  strongest  for slightly  off-critical
compositions $m_0^{-1}\sim \of{c_a-c_{a,c}^{(s)}} \lesssim 0$ the
binodals (and spinodals)  are shifted to smaller
values of $\zeta$ upon decreasing \abs{m_0}  down
to $m_0\simeq-20$. With a further decrease of \abs{m_0} the system
moves too far away from the critical point of the solvent so that the
CCF weakens and the spinodals  shift  again to larger values
of $\zeta$.

The critical value $\eta_c$ of the packing fraction is rather small,
i.e., $\eta_c\approx 0.07$ (see Fig.~\ref{fig:spinodals}), because 
the effective hard sphere diameter $\sigma$ which results from the 
soft repulsion $\pot[rep]$ is larger than $2R$.
In terms of $\eta_{\sigma}=(\sigma/(2R))^3\eta$
(defined after \eref{eq:febulk_pyrpa}) the critical value assumes
its RPA-value $\eta_{\sigma,c}^{\of{RPA}}=0.129$.
Furthermore, the binodals shown in Fig.~\ref{fig:spinodals} are
rather flat compared with, e.g., the ones for hard
spheres interacting via a short-ranged, attractive temperature
independent potential (described also by \eref{eq:febulk_pyrpa}).
In the present system, the  deviation from the critical temperature 
$T_c$ which leads to a range of $\eta$
for the coexisting phases as large as shown in Fig.~\ref{fig:spinodals},
is about $1$\textperthousand, whereas for a system of hard spheres with an
attraction the corresponding temperature deviation  is a few percent.
For smaller values of $\zeta_0$ the binodals are flatter and the differences
of the critical temperatures for different values of $m_0$ are 
smaller (see Fig.~\ref{fig:spinodals}(a)). 
The requirement $\eta^{-1/3}>\of{4\pi/3}^{1/3}\of{2+\zeta/s}$ 
(see Subsect.~\ref{subsec:crit_ccf}) concerning the validity
of the pairwise approximation for the CCFs 
is fulfilled for the whole range of values 
for $\eta$ and $\zeta$ shown in 
\fref{fig:spinodals}; 
e.g., for $s=10$ and $\zeta=4$ the above condition is $\eta<0.3$.

\begin{figure}
	\includegraphics{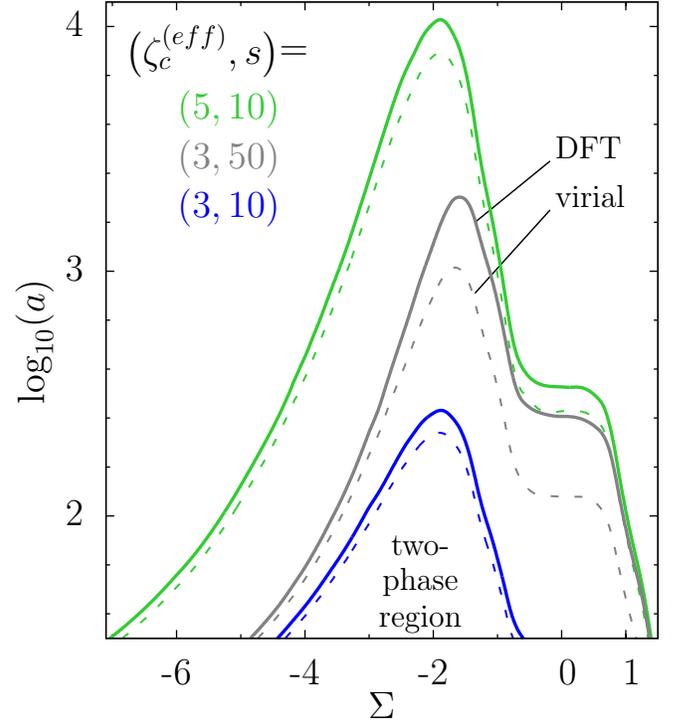}
	\caption{
	Attraction due to  CCFs  can induce a ``liquid''-``gas'' phase
	separation of the colloids. Within the effective approach
	described by \eref{eq:coll_potential} the dominant temperature
	dependence is that of the Casimir scaling function encoded in
	$\zeta=\kappa \xi\of{t}\sim\abs{t}^{-\nu}$. 
	The full curves correspond to DFT (\eref{eq:dft_rpa}) and show 
	for which values of $\Sigma$ and $a$ corresponding to the solvent
	composition and the repulsive strength of the background potential, 
	respectively, a given temperature (in terms of $\zeta$) is the
	critical one $T^{(eff)}_{c}$ or $\zeta^{(eff)}_{c}$.
	The contour lines  $a\of{\Sigma;s;\zeta_c^{\of{eff}}}$ depend 
	parametrically on $s=\kappa R$ where $\kappa^{-1}$ is the range 
	of the background repulsion and $R$ is the radius of the colloids.
	Systems, with their parameters being located in the area below
	the contour line $\zeta^{(eff)}_{c}$, are
	phase separated at this temperature. This two-phase region
	expands to larger values of $a$ with increasing $\zeta$. 
	For a fixed value $\zeta^{(eff)}_{c}$, 
	$a\of{\Sigma;s;\zeta_c^{\of{eff}}}$ is
	largest for compositions slightly poor in the component preferred 
	by  colloids, i.e., $\Sigma\lesssim0$.
	In addition, the corresponding  contour lines based on 
	the VLNF criterion  
	$\bsec^*\of{\zeta=\tilde\zeta^{(eff)}_{c},\Sigma,a;s}=\bsec[,c]^*$
	for the reduced second {\it{}virial} coefficient 
	(\eref{eq:stickyparam}) are plotted  as dashed lines.
	It is satisfactory to see that both approaches provide 
	comparable results.}
	\label{fig:tau_diagram}
\end{figure}
The sketch of the phase diagram in \fref{fig:phases_sketch} and the 
calculated coexistence curves in \fref{fig:spinodals} correspond to 
using the background potential with fixed parameters $s$ and $a$. 
In order to investigate the dependence of the locus of the phase 
separation on properties of the background potential, in the following 
we shall discuss how the critical temperature  
$T^{(eff)}_{c}=T^{(eff)}_{c}\of{\Sigma,a,s,\zeta_0,\tcb,\pm}$ of the 
effective colloidal system varies as function of its arguments 
($\pm$ is taken to be $-$ and $+$ for a lower and 
an upper critical point, respectively). 
According to the discussion above, to this end 
it is sufficient to determine $\zeta^{(eff)}_{c}\of{\Sigma,a,s}$. 
In \fref{fig:tau_diagram} for fixed values of $s$ and for specific 
values of $\zeta^{(eff)}_{c}$ we show the contour lines
$a=a\of{\Sigma;s;\zeta^{(eff)}_{c}}$.
That is, for a colloidal system with a  background potential 
characterized by certain values of  $a$ and $s$, from the plot in 
\fref{fig:tau_diagram} one can read off the value of the difference 
of the chemical potentials of the two components of the solvent (which 
is related to $\Sigma$) for which $T^{(eff)}_{c}$
has a prescribed value. The full curves are calculated within the 
density functional approach (\eref{eq:dft_rpa})  and are compared 
with the ones corresponding to the simple prediction
(Eqs.~\eqref{eq:secvirial} and \eqref{eq:coll_potential}) 
$\bsec^*\of{\zeta=\tilde\zeta^{(eff)}_{c},\Sigma,a;s}=\bsec[,c]^*=-1.212$,
which also yields a relation 
$\tilde\zeta^{(eff)}_{c}=\tilde\zeta^{(eff)}_{c}\of{\Sigma,a;s}$. 
Both approaches differ only slightly. For given values of  
$\zeta_c^{\of{eff}}$ and $\Sigma$ the VLNF conjecture  predicts a 
smaller critical value of $a$ than obtained  from DFT. The largest 
deviations occur for slightly negative values of $\Sigma$, 
for  which the CCFs are  more attractive. 
For all values of the parameters shown in \fref{fig:tau_diagram}  
(i.e., $a \gtrsim 30$), 
at small distances $D$ the corresponding potentials exhibit repulsive 
barriers  $\gtrsim 30 k_{B} T$. For smaller values of $a$, i.e., 
$a\lesssim 15$, this potential barrier disappears and the colloidal
particles may form aggregates. This is discussed in the following
paper \cite{MMD_part2}; see Figs.~1 and 2 therein.

The contour lines $a=a\of{\Sigma;s;\zeta^{(eff)}_{c}}$ attain their
largest values for slightly negative values of $\Sigma$, i.e., for 
$\Sigma\simeq -2$. This again reflects the asymmetry of the strength 
of the CCFs with respect to the bulk ordering field \hb{} in the 
presence of symmetry breaking surface fields. 
In \fref{fig:tau_diagram}, for all 
parameter pairs \of{a,\Sigma}  below each contour line, characterized 
by $\zeta^{(eff)}_{c}$, the corresponding system at the temperature 
belonging to the value $\zeta^{(eff)}_{c}$ is phase separated. 
This two-phase region widens with increasing values of $\zeta$ and 
it significantly expands to larger values of $a$, which is in line 
with the fact that the CCFs become stronger upon approaching \tcb{}.
For $s=10$, upon  increasing $\zeta^{(eff)}_{c}$ from $3$
to $5$ the largest critical value of $a$ increases by two orders of
magnitude  (compare the corresponding blue and green curves in
\fref{fig:tau_diagram}).  
The two-phase region can be markedly increased upon increasing $s$ 
(compare the blue and gray curves in \fref{fig:tau_diagram} 
corresponding to $\zeta_c^{\of{eff}}=3$ with $s=10$ and $s=50$, 
respectively) because for the phase separation the strength of the 
attraction is important; due to $\pot\sim s$ 
(\eref{eq:coll_potential}) the latter can be easily 
tuned by varying $s$.

For the experimental studies reported in
Refs.~\onlinecite{Kline-et:1994,Jayalakshmi-et:1997} colloidal mixtures 
consisting of silica spheres, water, and 2-butoxyethanol or lutidine
have been used; with a radius $R\approx 11nm$ the colloidal particles 
have been rather small. Thus, a priori, the effective approach,
developed in Subsect.~\ref{sec:colloids} and discussed in 
Subsect.~\ref{ssc:eff}, is not expected to apply.
Nonetheless, one can try to do so
and estimate the critical temperature for solutions containing such
small particles  by using the general scaling arguments presented 
in Subsect.~\ref{ssc:sc} and in Ref.~\onlinecite{note_CP_scaling}. 
At the critical temperature, the dimensionless scaling variable 
$w^{(3)}= w_0^{(3)}\abs{t}^{\nu}R$ associated with
the radius of the colloid takes on a certain value $w_c^{(3)}$ with
$w_c^{(3)}/w_0^{(3)}=\abs{\of{\tcb-T_c(\rho)}/\tcb}^{\nu}R\equiv \abs{t_c}^{\nu}R$
(where $w_0^{(3)}$  is a non-universal amplitude). Using our effective
approach we can calculate $w_c^{(3)}/w_0^{(3)}$ within mean-field theory
for parameters corresponding to the solvents used in those experiments,
and demand \cite{note_w3c} that this ratio applies approximately to the
experiments reported in Refs.~\onlinecite{Kline-et:1994,Jayalakshmi-et:1997}. 
For the solvents used, the relevant parameters are
$\xi_0\approx 0.25 nm$  and  $\zeta_0\sim0.03$. 
Choosing $s=10$ (see Subsect.~\ref{ssc:res_range} and \fref{fig:spinodals}(a))
renders $R=\xi_0 s/\zeta_0 \approx 85nm$. This value of $R$ is
sufficiently large so that the aforementioned effective approach can 
be considered to be reliable.  From \fref{fig:spinodals}(a)  we can 
read off the corresponding shift
$\abs{t_c} \sim 0.001$ which leads to  $w_c^{(3)}/w_0^{(3)}\simeq 1.094nm$.
By using this latter mean-field estimate, 
for the experimental value $R\approx 11 nm$ of the radius of the
colloidal particles one thus obtains the shift
$\abs{t_c}=\of{w_0^{(3)}R/w_c^{(3)}}^{-1/\nu}\simeq0.234$.
With the critical temperature $\tcb\approx320K$ of the solvent
\cite{Kline-et:1994,Jayalakshmi-et:1997}, this leads to
$T_c\of{\rho}= \of{1-t_c}\tcb\approx 310 K$.
Indeed, for such a temperature phase separation in the ternary mixture
has been observed (although even lower critical temperatures
$T_c \approx 300K$ have been  reported 
\cite{Kline-et:1994,Jayalakshmi-et:1997}). 
As stated above such an agreement for even small colloids could not 
have been anticipated from the outset.

\section{Summary \label{sec:summary}}

We have studied the collective behavior of  monodisperse colloidal 
suspensions with  near-critical solvents. Colloidal particles acting 
as cavities set the boundary conditions for the fluctuating 
order parameter (OP) of the solvent and  perturb  the OP field 
on the length scale of the bulk correlation length 
$\xi\of{t=\of{T-\tcb}/\tcb\to 0}\sim \abs{t}^{-\nu}$, which diverges 
upon approaching the critical temperature \tcb{}; $\nu$ is a critical 
bulk exponent. These modifications of the OP and restrictions of its 
fluctuation spectrum result in an effective force acting between the 
colloids, known as the critical Casimir force (CCF). For equal boundary 
conditions (BCs), i.e., equal surface properties of all colloids, the 
CCF is attractive. The CCF depends on the configuration of all colloidal 
particles and thus is in general non-additive. We have obtained our 
results for an effective one-component system of colloids interacting 
via an effective pair potential. 
The regular background potential 
has been taken to be a soft repulsion acting on a length scale 
$\kappa^{-1}$ and with strength $A$. We have considered suspensions 
with medium values of the colloidal number density 
$\rho$, for which the approximation of pairwise additive CCFs is valid. 
The scaling function of the critical Casimir potential has been taken in 
accordance with the Derjaguin approximation 
(\eref{eq:sfpot_derjaguin}), within which the pair potential between 
two spheres is expressed in terms of the scaling function of the CCF in a slab. 
The variation of the scaling function in a slab with the scaling variable 
$\Sigma$ associated with the bulk ordering field $\hb$ (see 
\eref{eq:coll_potential0}) in spatial dimension $d=3$ has been 
approximated based on Monte Carlo simulation data and on mean-field
theory results (\eref{eq:sfp_h_approx}). On this basis we have 
obtained the following main results: 
\begin{enumerate}
\item
	In \fref{fig:phases_sketch} we have illustrated
	the phase behavior of a ternary mixture consisting of colloidal 
	particles  and a binary solvent which exhibits a miscibility gap.
	We have discussed in detail the relationship
	between this full description and the one which is based on
	an effective, one-component colloidal fluid
	(\fref{fig:phases_sketch}).

\item
	For certain regions of the three-dimensional thermodynamic 
	space \of{T,c_a,\rho} of a ternary mixture, the effective
	one-component model is applicable for obtaining the onset
        of phase separation. For particles 
	interacting via the effective pair potential  given in
	\eref{eq:coll_potential} we have calculated the phase coexistence
	curve and the spinodal (i.e., the  loci
	where within mean-field theory the isothermal compressibility
	$\chi_T$ diverges)  both within density functional
	theory (DFT, \eref{eq:dft_rpa}) and the integral equation
	approach. In \fref{fig:spinodals} the coexistence curves for
	various values of $c_a-c_{a,c}^{(s)}\sim m_0^{-1}$ are shown.
      	The spinodals as obtained by using the DFT approach are narrower 
	and are located at smaller values of 
        $\zeta=\kappa\xi\sim\abs{t}^{-\nu}$ than the ones
	obtained from the integral equation approach.
\item
	In \fref{fig:tau_diagram} the dependence of the critical
	temperature of the effective one-component
	system (expressed in terms of $\zeta^{(eff)}_{c}$) 
	on the parameters $a=A/s$, $\Sigma$, and
        $s=\kappa R$ is shown.
	For a given value of $a$, $\zeta^{(eff)}_{c}$ is smallest
	for slightly negative values of $\Sigma$.
	The critical temperature has been calculated within DFT and
	compared with the simple prediction $\bsec^{*}=\bsec[,c]^{*}$
	as suggested by Vliegenthart and Lekkerkerker
	\cite{Vliegenthart-et:2000} and Noro and Frenkel
	\cite{Noro-et:2000}; $\bsec^*$ is the reduced second virial
	coefficient (Eqs.~\eqref{eq:secvirial} and
	\eqref{eq:stickyparam}) and $\bsec[,c]^{*}$ is the critical 
	value of Baxter's model. Both approaches yield good agreement.  
\end{enumerate}

To conclude, our results show that the CCF, which can be easily 
controlled by temperature $T$ and the strength and range of which 
can be varied by changing the bulk ordering field \hb{}, can induce  
phase separation into a colloidal-poor and a colloidal-rich phase.
Within the approach  of using an  effective potential,
we have  identified the ranges of values for the background repulsive 
potential and the values of the scaling variables associated with the 
critical solvent, for which the colloids  phase separate. 
We have used the approach developed here in order to 
study also the stability of colloidal suspensions 
in near-critical binary solvents \cite{MMD_part2}.

Concerning further research, more complex suspensions
could be considered, for example a mixture of colloids with different
adsorption preferences for the solvent particles.
According to Gibbs' phase rule, for such a four-component system 
(a binary mixture  of colloids in a binary solvent) at constant
pressure a  two-dimensional manifold of critical points embedded 
in a three-dimensional manifold of coexisting states in the (then 
four-dimensional) thermodynamic space can be expected.
Recently, such a mixture was studied  experimentally 
\cite{Zvyagolskaya-et:2011}. In this study, concerning the effective 
interaction potential of the CCFs between the colloids the attractive 
\of{+,+} BCs and the repulsive \of{+,-} BCs were realized for colloidal
particles of the same kind and for particles of different kind, respectively. 
In line with an effective DFT, upon approaching the critical temperature of 
the solvent there are indications that these two kinds of 
particles phase separate.

In order to study the phase behavior of the actual three- or 
four-component mixture for the full range of all thermodynamic 
variables $T$, $c_a$, and $\rho$, an analysis beyond the effective
approach is needed.

\begin{acknowledgments}

We thank R. Evans for his generous interest in our work and 
the stimulating and enlightening discussions we enjoyed to 
have with him. 
\end{acknowledgments}

\appendix
\section{Off-critical mixtures \label{app:1}}
The Hamiltonian given in \eref{eq:lgh} depends on the reduced temperature 
$t$ and the bulk field \hb{}, which for a binary liquid mixture is
proportional to the deviation of the difference of the chemical potentials 
of the two components $a$ and $b$ from its critical value, i.e., 
$\hb\sim \of{\mu_a -\mu_b} -\of{\mu_{a,c} -\mu_{b,c}}$ for
a fixed pressure $\sim \mu_a +\mu_b$. 
Accordingly, for fixed \hb{} the OP varies upon changing $t$. However, in
most experimental realizations  $t$ is varied at fixed concentrations 
$c_i$, $i\in\set{a,b}$, and thus the OP is kept constant.
The OP $\phi$, which is not uniquely defined, is related to the ordering 
field $\hb$  by the equation of state (EOS) which  for a critical system
takes the scaling form \cite{Pelissetto-et:2002}
\begin{equation}
\label{eq:eos_critical}
  \hb=\mathcal{D}\sgn(\phi)\abs{\phi}^{\delta} \sfeos\of{t \abs{\mathcal{B}/\phi}^{1/\beta} },
\end{equation}
or equivalently
\begin{equation}
\label{eq:eos_critical_op}
	\Op =\sgn\of{\hb}\mathcal{B}\abs{t}^{\beta}
	G\of{\mathcal{B}^{1/\beta}\mathcal{D}^{1/\of{\delta\beta}} t
	  \abs{\hb}^{-1/\of{\delta\beta}}}
\end{equation}
with universal scaling functions $\sfeos\of{\hat{X}}$ and $G\of{Y}$
with $G\of{Y\to 0}=\abs{Y}^{-\beta}$. 
$\mathcal{D}$ and $\mathcal{B}$ are non-universal amplitudes which 
depend on the definition of $\phi$ such that on the coexistence curve 
the bulk  OP follows 
$\phi_b\of{t\to0^{-},\hb=0}=\mathcal{B}\abs{t}^{\beta}$.
$\mathcal{D}$, $\mathcal{B}$, and the correlation length amplitudes 
$\xi_0^{\pm,h}$ (defined after \eref{eq:ccf_spsp_scaling_org}) are
related to each other by universal amplitude ratios such that only 
two of them are independent \cite{Pelissetto-et:2002,colloids1b}.
In the lowest order in its argument   
\begin{equation}
	\label{eq:x}
	\hat{X}= t\abs{\mathcal{B}/\phi}^{1/\beta},
\end{equation}
the universal scaling  function \sfeos{} has the functional form
$\sfeos\of{\hat X}=1+\hat{X}$, 
which captures the crossover between the critical behavior
at $t=0$ and at  $\hb=0$, respectively \cite{Pelissetto-et:2002}.
At the coexistence curve one has $t<0$ and $\hb=0$;
therefore $\hat X=-1$ in agreement with 
$\phi=\mathcal{B}\abs{t}^{\beta}$.
Along the critical isotherm $t=0$ one has $\hat{X}=0$ and thus 
$\phi=\sgn\of{\hb}\abs{\hb/\mathcal{D}}^{1/\delta}$.
For our purpose of fluid systems exposed to surfaces the sign of $\phi$ 
matters and the appropriate scaling variable is 
$X=\sgn\of{\phi}\abs{t}\abs{\mathcal{B}/\phi}^{1/\beta}$.
In terms of the scaling variables 
$\Sigma=\sgn\of{\hb}\xi/\xi^{\of{h}} =
      \sgn\of{\hb}\of{\xi_0^{+}/\xi_0^{\of{h}}}
      {\abs{\hb}^{\nu/\of{\beta\delta}}\abs{t}^{-\nu}}$ 
and $X$, \eref{eq:eos_critical} 
takes the scaling form
\begin{equation}
\begin{split}
\label{eq:eos_critical_scaling}
 \sgn\of{\Sigma}  \abs{\Sigma}^{\beta\delta/\nu} & =  \\
    & \of{R_{\chi}\delta/Q_2}^{\delta/\of{\delta-1}}
    \of{Q_{\xi}^{+}/Q_{\xi}^{c}}^{\beta\delta/\nu}  \\ & \times
    \sgn\of{X} \abs{X}^{-\beta\delta} \sfeos[\pm]\of{\abs{X}},
\end{split}
\end{equation}
where $\sfeos[\pm]\of{\abs{X}}=1\,\pm\,\abs{X}$ and
$\pm$ refers to the sign of $t$. 
$R_{\chi}$, $Q_2$, $Q_{\xi}^{+}$, and $Q_{\xi}^{c}$ 
are universal amplitude ratios \cite{Pelissetto-et:2002}.
We shall use these expressions in order to calculate the variation 
of several experimentally accessible
quantities along experimentally realizable thermodynamic paths.




\begin{thebibliography}{}


\bibitem{Likos:2001} C. N. Likos, Phys. Rep. {\bf 348}, 267 (2001).

\bibitem{Asakura-et:1954} 
S. Asakura and F. Oosawa, J. Chem. Phys. {\bf 22}, 1255 (1954);
concerning systems in which adding small particles weakens the net 
attraction between large particles see, e.g., 
D. J. Ashton, J. Liu, E. Luijten, and N. B. Wilding, 
J. Chem. Phys. {\bf 133}, 194102 (2010).

\bibitem{Dijkstra-et:1999}
M. Dijkstra, J. M. Brader, and R. Evans, J. Phys.: Condens. Matter {\bf 11}, 10079 (1999).

\bibitem{Buzzaccaro-et:2007}
S. Buzzaccaro, R. Rusconi, and R. Piazza, Phys. Rev. Lett. {\bf 99}, 098301 (2007).

\bibitem{Evans:1990} R. Evans, J. Phys.: Condens. Matter {\bf 2}, 8989 (1990).

\bibitem{Fisher-et1978}
M. E. Fisher and P. G. de Gennes, C. R. Acad. Sci., Paris, Ser. B {\bf 287}, 207 (1978).

\bibitem{Hertlein-et2008} 
C. Hertlein, L. Helden, A. Gambassi, S. Dietrich, and C. Bechinger,
Nature {\bf 451}, 172 (2008).

\bibitem{Gambassi-et:2009}
A. Gambassi, A. Macio\l ek, C. Hertlein, U. Nellen, L. Helden, C. Bechinger, and S. Dietrich,
Phys. Rev. E {\bf 80}, 061143 (2009).

\bibitem{Nellen-et:2009}
U. Nellen, L. Helden, and C. Bechinger, EPL {\bf 88}, 26001 (2009).



\bibitem{colloids2}
T. W. Burkhardt and E. Eisenriegler,
Phys. Rev. Lett. {\bf 74}, 3189 (1995).

\bibitem{dantchev_dietrich}
D. Dantchev, F. Schlesener, and S. Dietrich, 
Phys. Rev. E {\bf 76}, 011121 (2007).

\bibitem{FSS} (a) M. N. Barber in {\it Phase Transitions and Critical Phenomena},
edited by C. Domb and J. L. Lebowitz (Academic, New York, 1983), Vol. 8, p. 149; 
(b) V. Privman, in 
{\it Finite Size Scaling and Numerical Simulation of  Statistical Systems}, 
edited by V. Privman (World Scientific, Singapore, 1990), p. 1.

\bibitem{Diehl:1986} H. W. Diehl in {\it Phase Transitions and Critical Phenomena},
edited by C. Domb and J. L. Lebowitz (Academic, New York, 1986), Vol. 10,
p. 76.

\bibitem{Krech:1990:0}
M. Krech, {\it Casimir Effect in Critical Systems} (World Scientific,
Singapore, 1994); 
J. Phys.: Condens. Matter {\bf 11}, R391 (1999).


\bibitem{Dbook}
G. Brankov, N. S. Tonchev, and D. M. Danchev,
{\it Theory of Critical Phenomena in Finite-Size Systems}
(World Scientific, Singapore, 2000). 


\bibitem{gambassi}  More recent results for critical Casimir forces are summarized
 in A. Gambassi, J. Phys.: Conf. Ser. {\bf 161}, 012037 (2009).

\bibitem{colloids1a}
A. Hanke, F. Schlesener, E. Eisenriegler, and S. Dietrich,
Phys. Rev. Lett. {\bf 81}, 1885 (1998).

\bibitem{colloids1b}
F. Schlesener, A. Hanke, and S. Dietrich, 
J. Stat. Phys. {\bf 110}, 981 (2003).

\bibitem{khd-08}
S. Kondrat, L. Harnau, and S. Dietrich, J. Chem. Phys. {\bf 131},  183901 (2009).

\bibitem{krech}
M. Krech, Phys. Rev. E {\bf 56}, 1642 (1997).

\bibitem{upton}
(a) Z. Borjan and P. J. Upton,  Phys. Rev. Lett. {\bf 81}, 4911 (1998);
(b) {\it ibid} {\bf 101}, 125702 (2008).

\bibitem{FdeG_loc_fun} 
(a) M. E. Fisher and P. J. Upton, Phys. Rev. Lett. {\bf 65}, 2402 (1992);
(b) {\it ibid} {\bf 65}, 3405 (1992).


\bibitem{vas}
O. Vasilyev, A. Gambassi, A. Macio{\l}ek, and S. Dietrich, 
EPL {\bf 80}, 60009 (2007); 
O. Vasilyev, A. Gambassi, A. Macio{\l}ek, and S. Dietrich, 
Phys. Rev. E {\bf 79}, 041142 (2009).

\bibitem{Hasenbusch}  
M. Hasenbusch, Phys. Rev. B {\bf 82}, 104425 (2010).

\bibitem{Hasenbusch-cross}
 M. Hasenbusch, Phys. Rev. B {\bf 83}, 134425 (2011).

\bibitem{vas-cross} 
O. Vasilyev, A. Macio\l ek, and S. Dietrich,
Phys. Rev. E {\bf 84}, 041605  (2011).

\bibitem{Buzzaccaro-et:2010} 
(a) S. Buzzaccaro, J. Colombo, A. Parola, and R. Piazza, 
Phys. Rev. Lett. {\bf 105}, 198301 (2010); 
(b) R. Piazza, S. Buzzaccaro, A. Parola, and J. Colombo, 
J. Phys.: Condens. Matter {\bf 23}, 194114 (2011). 

\bibitem{pershan}
M. Fukuto, Y. F. Yano, and P. S. Pershan, 
Phys. Rev. Lett. {\bf 94}, 135702 (2005).

\bibitem{rafai}
S. Rafa{\"\i}, D. Bonn, and J. Meunier, Physica A {\bf 386}, 31 (2007).

\bibitem{Derjaguin:1934} 
B. Derjaguin, Kolloid Zeitschrift {\bf 69}, 155 (1934).

\bibitem{Troendle-et:2010}
M. Tr{\"o}ndle, S. Kondrat, A. Gambassi, L. Harnau, and S. Dietrich,
EPL {\bf 88}, 40004 (2009);
M. Tr\"{o}ndle, S. Kondrat, A. Gambassi,  L. Harnau, and S. Dietrich,
J. Chem. Phys. {\bf 133}, 074702 (2010);
M. Tr{\"o}ndle, O. Zvyagolskaya, A. Gambassi, D. Vogt, L. Harnau, C. Bechinger, and S. Dietrich,
Mol. Phys. {\bf 109}, 1169 (2011).

\bibitem{Kline-et:1994} 
S. R. Kline and E. W. Kaler, Langmuir {\bf 10}, 412 (1994).

\bibitem{Jayalakshmi-et:1997}
Y. Jayalakshmi and E. W. Kaler, Phys. Rev. Lett. {\bf 78}, 1379 (1997).

\bibitem{Koehler-et:1997} 
R. D. Koehler and E. W. Kaler, Langmuir {\bf 13}, 2463 (1997).

\bibitem{Sluckin:1990} 
T. J. Sluckin, Phys. Rev. A {\bf 41}, 960 (1990).

\bibitem{Loewen:1995} H. L{\"o}wen, Phys. Rev. Lett. {\bf 74}, 1028 (1995).

\bibitem{Netz:1996} R. R. Netz, Phys. Rev. Lett. {\bf 76}, 3646 (1996).

\bibitem{Barrat-et:2003}
J.-L. Barrat and J.-P. Hansen, 
{\it Basic concepts for simple and complex liquids} 
(Cambridge University Press, Cambridge, 2003).

\bibitem{Russel-et:1989}
W. B. Russel, D. A. Saville, W. R.  Schowalter, 
{\it Colloidal Dispersions} (Cambridge University Press, 1989).

\bibitem{Pelissetto-et:2002}
A. Pelissetto  and E. Vicari, Phys. Rep. {\bf 368}, 549 (2002).

\bibitem{Hansen-et:1976}
J. P. Hansen and I. R. McDonald, {\it Theory of Simple Liquids} 
(Academic, London, 1986). 

\bibitem{Caccamo:1996}
C. Caccamo, Phys. Rep. {\bf 274}, 1 (1996).


\bibitem{Binder-et:1978}
K. Binder, C. Billotet and P. Mirold, Z. Physik B {\bf 30}, 183 (1978).


\bibitem{Ornstein-et:1914} 
L. S. Ornstein,  and F. Zernike, Proc. Acad. Sci. Amsterdam {\bf 17}, 793 (1914).

\bibitem{Gillian:1979}
M. J. Gillan, Mol. Phys. {\bf 38}, 1781 (1979).


\bibitem{Archer-et:2007}
A. J. Archer and N. B. Wilding, Phys. Rev. E {\bf 76}, 031501 (2007).

\bibitem{Kurnaz}
(a) M. L. Kurnaz and J. V. Maher, Phys. Rev. E  {\bf 51}, 5916 (1995); 
(b) M. L. Kurnaz and J. V. Maher, Phys. Rev. E {\bf 55}, 572 (1997).

\bibitem{Vliegenthart-et:2000}
G. A. Vliegenthart and H. N. W. Lekkerkerker, 
J. Chem. Phys. {\bf 112}, 5364 (2000).

\bibitem{Noro-et:2000}
M. G. Noro and D. Frenkel, J. Chem. Phys. {\bf 113}, 2941 (2000).

\bibitem{thN} 
(a) J. Largo and N. B. Wilding, Phys. Rev. E {\bf 73}, 036115 (2006);
(b) G. Foffi and F. Sciortino, Phys. Rev. E {\bf 74}, 050401(R) (2006);
(c) P. Orea and Y. Duda, J. Chem. Phys. {\bf 128}, 134508 (2008);
(d) D. Gazzillo, J. Chem. Phys. {\bf 134}, 124504 (2011).


\bibitem{Baxter:1968} 
R. J. Baxter, J. Chem. Phys. {\bf 49}, 2770 (1968).

\bibitem{Miller-et:2003}
M. A. Miller  and D. Frenkel, Phys. Rev. Lett. {\bf 90}, 135702 (2003); 
M. A. Miller  and D. Frenkel, J. Chem. Phys. {\bf 121}, 535 (2004).

\bibitem{Weeks-et:1971}
J. D. Weeks, D. Chandler,  and H. C. Andersen,
J. Chem. Phys. {\bf 54}, 5237 (1971).

\bibitem{Andersen-et:1971}
H. C. Andersen, J. D.  Weeks, and  D. Chandler, 
Phys.  Rev. A {\bf 4}, 1597 (1971).

\bibitem{Evans:1979} 
R. Evans, Adv. Phys. {\bf 28}, 143 (1979).

\bibitem{DFTcolloids}  see, e.g., 
  J. M. Brader, R. Evans, and M. Schmidt, Mol. Phys. {\bf 101}, 3349 (2003),
  and references therein.

\bibitem{Rosenfeld:1989} Y. Rosenfeld, Phys. Rev. Lett. {\bf 63}, 980 (1989).

\bibitem{Roth-et:2002} R. Roth, R. Evans, A. Lang,  and G. Kahl, J. Phys.: Condens. Matter {\bf 14}, 12063 (2002).
 
\bibitem{HansenGoos-et} H. Hansen-Goos and R. Roth, J. Phys.: Condens. Matter {\bf 18}, 8413 (2006).


\bibitem{Abraham-et:2010}
D. B. Abraham and A. Macio\l ek, Phys. Rev. Lett. {\bf 105}, 055701 (2010).

\bibitem{Nowakowski-et:2009} P. Nowakowski and M. Napiorkowski, 
J. Phys. A {\bf 42}, 475005 (2009).

\bibitem{Mohry-et:2010}
T. F. Mohry, A. Macio{\l}ek, and S. Dietrich, Phys. Rev. E {\bf 81}, 061117 (2010).


\bibitem{Beysens-et:1985}
D. Beysens and D. Est\`eve, Phys. Rev. Lett. {\bf 54}, 2123 (1985).


\bibitem{Nellen_hdependence}
U. Nellen, doctoral thesis, University of Stuttgart (2011).

\bibitem{tobepublished} T. F. Mohry, A. Macio{\l}ek, and S. Dietrich, unpublished.

\bibitem{Eisenriegler-et:1994}
E. Eisenriegler and M. Stapper, Phys. Rev. B {\bf 50}, 10009 (1994).

\bibitem{Nellen-et:2011}
U. Nellen, J.  Dietrich, L. Helden, S. Chodankar, K. Nygard, J. F. van der Veen, and C. Bechinger, 
Soft Matter {\bf 7}, 5360 (2011).

\bibitem{Ciach-et:2010}
A. Ciach and A. Macio{\l}ek, Phys. Rev. E {\bf 81}, 041127 (2010);
F. Pousaneh and A. Ciach, J. Phys.: Condens. Matt. {\bf 23}, 412101 (2011).

\bibitem{Bier-et:2010}
M. Bier, A. Gambassi, M. Oettel, and S. Dietrich, 
EPL {\bf 95}, 60001 (2011).

\bibitem{Jayalakshmi-et:1994}
Y. Jayalakshmi, J. S. Van Duijneveldt, and D. Beysens, J. Chem. Phys. {\bf 100}, 604 (1994).

\bibitem{Handschy-et:1980}
M. A. Handschy, R. C. Mockler, and W. J. O'Sullivan, 
Chem. Phys. Lett. {\bf 76}, 172 (1980).

\bibitem{Jacobs-et:1977}
D. T. Jacobs, D. J. Anthony, R. C. Mockler, and W. J. O'Sullivan,
Chem. Phys. {\bf 20}, 219 (1977).

\bibitem{WL-note}
For the mixture of  2,6-lutidine and water the
amplitude ${\mathcal B}$ of the bulk OP 
$\phi\of{t\to 0^{-}}={\mathcal B}\abs{t}^{\beta}$
can be estimated from the refractive indices $n_{1,2}$  of the two 
coexisting phases $1$ and $2$ near the lower critical point 
\cite{Handschy-et:1980}. The refractive index difference 
$\Delta n=n_{1}-n_{2}$ 
is related to the difference in the volume fraction 
$\Delta \Phi_a=\Phi_a^{(1)}-\Phi_a^{(2)}$ 
of the component $a$ in the two phases according to 
\cite{Handschy-et:1980} $\Delta \Phi_a=k(T)\Delta n$. 
The coefficient $k(T)$ can be expressed \cite{Jacobs-et:1977} in terms 
of the refractive indices $n_{1,2}$ and the refractive indices 
$n_{a,b}$ of the pure components $a$ and $b$:
%
$k=3 \of{n_1 +n_2 } \fd{\of{A_a-A_b}\of{n_1^2+2}\of{n_2^2+2}}^{-1}$, 
%
where $A_{a,b}=\of{n_{a,b}^2-1}/\of{n_{a,b}^2+2}$. 
The refractive  indices of pure  water and 
pure  2,6-lutidine are \cite{Jayalakshmi-et:1994} 
$n_{W}=1.33$  and $n_L=1.49$,  respectively.
For the lutidine-water mixture the resulting value of $k=k(T)$ varies 
by less than $0.4\%$ within the reported temperature range
\cite{Handschy-et:1980} $T-\tcb<15K$. Thus to a good approximation one can take 
$k\of{T} \simeq k_0 \simeq 6.35$ independent of $T$.
According to Ref.~\onlinecite{Handschy-et:1980} for $T-\tcb{}< 0.5K$ the 
two-phase coexistence in terms of the refractive index is well 
described by the power law
$\Delta n=A \of{T-\tcb}^{\beta}$  with 
$\tcb =\of{307.258\pm 0.001} K$,
$A=\of{0.0471\pm 0.0001}K^{-\beta}$, and ${\beta}=0.338\pm 0.003$.
Thus for 
$\Delta  \Phi \simeq { B_{\Phi}}\abs{t}^{\beta} $
we obtain the amplitude
${ B_{\Phi}}= k_0 \of{\tcb}^{\beta} A \simeq 2.073$.
%
The mass fraction \m[a] in terms of the volume fraction $\Phi_{a}$ is given by 
$\m[a]=\rho_{m,a}\Phi_a\fd{\rho_{m,a}\Phi_a+\rho_{m,b}\Phi_b}^{-1}
=\Omega\Phi_a\fd{\of{\Omega-1}\Phi_a+1}^{-1}$,
where $\rho_{m,a}$ ($\rho_{m,b}$) is the mass density of the component $a$ ($b$) 
and  $\Omega=\rho_{m,a}/\rho_{m,b}$.
Accordingly 
$\Delta\m[a]\of{t\to 0^-}=\m[a]^{(1)}-\m[a]^{(2)}={B}_{\m}\abs{t}^{\beta}$,
where the amplitude is given by
${B}_{\m}=\Omega\fd{\of{\Omega-1}\Phi_{a,c}+1}^{-2}{B}_{\Phi}
=\Omega\fd{1+\of{\Omega^{-1}-1}\m[a,c]}^{2}{B}_{\Phi}$;
$\m[a,c]$ is the critical mass fraction of the component $a$.
With the mass densities \cite{Jayalakshmi-et:1994} $\rho_{m,W}=0.995 g/cm^3$ 
and $\rho_{m,L}=0.911g/cm^3 $ of pure water and pure lutidine, respectively, 
and the critical mass fraction \cite{Beysens-et:1985}
$\m[L,c]\simeq 0.29$ we obtain ${B}_{\m}\simeq 2.001$.
%
For the symmetric coexistence curve, as assumed here, the OP defined 
by the mass fraction is
$\Op[\m]\of{t<0,\hb=0}= \m[a]^{\of{1}}-\m[a,c]=0.5\Delta\m[a]$,
rendering the value $\mathcal{B}_{\m}=0.5\times{B}_{\m}\approx 1.00$.

\bibitem{Gnan-et:2011} 
N. Gnan, E. Zaccarelli, P. Tartaglia, and F. Sciortino,  
Soft Matter {\bf 8}, 1991 (2012).


\bibitem{Andon-et:1952} 
R. J. L. Andon and J. D. Cox, J. Chem. Soc. {\bf 1952}, 4601 (1952);
J. D. Cox, J. Chem. Soc. {\bf 1952}, 4606 (1952).

\bibitem{Prafulla-et:1992}
B.V. Prafulla, T. Narayanan, and A. Kumar, Phys. Rev. A {\bf 46}, 7456 (1992).

\bibitem{terMixTheory}
D. Mukamel and M. Blume, Phys. Rev. A {\bf 10}, 610 (1974);
J. Sivardi{\`e}re and J. Lajzerowicz, Phys. Rev. A {\bf 11}, 2090 (1975).

\bibitem{Schmidt-et:2002}
M. Schmidt and A. R. Denton, Phys. Rev. E {\bf 65}, 021508 (2002).

\bibitem{Schmidt:2011}
M. Schmidt, J. Phys.: Condens. Matt. {\bf 23}, 415101 (2011).

\bibitem{Fisher-et:1981}
M. E. Fisher and H. Nakanishi, J. Chem. Phys. {\bf 75}, 5857 (1981).
 
\bibitem{note_CP_scaling}
Taking the scaling variable  $w^{\of{3}}\propto \abs{t}^{\nu }R$
into account, the proposed scaling of the free energy reads 
$\fe\of{T,\Delta\mu,\rho;R}\simeq 
       \abs{t}^{2-\alpha}K\of{w^{\of{1}},w^{\of{2}},w^{\of{3}}}$ 
and the critical values $w_c^{\of{1}}$ and $w_c^{\of{2}}$ depend on 
$w_c^{\of{3}}$. Thus one obtains 
$T_c\of{\rho}-\tcb{}\sim W_t\of{w_c^{\of{3}}} \rho^{1/\of{\nu d}}$ and 
$\hb[,c]\of{\rho}\sim W_h\of{w_c^{\of{3}}} \rho^{\Delta/\of{\nu d}}$  
with universal functions $W_{t,h}$.
Since $t_c\equiv \abs{T_c\of{\rho}-\tcb{}}/\tcb\to 0$ for $\rho \to 0$
it follows that $w_c^{\of{3}}\propto t_c^{\nu}R\to 0$.  In this limit the
functions $W_{t,h}$ are regular and reduce to $W_{t,h}\of{w=0}$. 
Accordingly, the leading behavior as given in the main text is recovered.

\bibitem{Zvyagolskaya-et:2011}
O. Zvyagolskaya, A. J. Archer, and C. Bechinger, EPL {\bf 96}, 28005 (2011).

\bibitem{gallagher:92}
(a) P. D. Gallagher and J. V. Maher, Phys. Rev. A {\bf 46}, 2012 (1992);
(b) P. D. Gallagher, M. L. Kurnaz, and J. V. Maher, Phys. Rev. A {\bf 46}, 7750 (1992).

\bibitem{grull:97}
(a) H. Gr{\"u}ll and D. Woermann, Ber. Bunsenges. Phys. Chem. {\bf 101}, 814 (1997);
(b) B. Rathke, H. Gr{\"u}ll, and D. Woermann, J. Colloid Interface Sci. {\bf 192}, 334 (1997).


\bibitem{note_col_in_solvent}
In Refs.~\onlinecite{gallagher:92,grull:97} also exceptions  
to the described behavior are discussed which, however, are
not relevant for the region of the thermodynamic space  considered  here. 
For high temperatures $T>\tcb$ (recall that \tcb{} is a lower
critical point) the colloids may populate the meniscus 
formed by  the two coexisting phases of the solvent, and for
compositions rather different from the critical one the colloids are
homogeneously distributed in both coexisting phases 
(as long as they are soluble at all).

\bibitem{MMD_part2}
T. F. Mohry, A. Macio{\l}ek, and S. Dietrich, following paper, 
preprint, arXiv:1201.5547.


\bibitem{note_w3c}
On one hand the value of $w_c^{(3)}/w_0^{(3)}$ depends on the 
specific components of the colloidal suspension via the non-universal 
amplitude $w_0^{(3)}$. On the other hand, according to 
Ref.~\onlinecite{note_CP_scaling}, it depends on $\rho$ (or equivalently on \hb, see 
\eref{eq:cp_shift}) via $w_c^{(3)}$ with $w_c^{(3)}$ as function of 
$w_c^{(2)}$ (or $w_c^{(1)}$). Thus, the value of $w_c^{(3)}/w_0^{(3)}$ 
varies along the line \cur[c] of critical points. While one knows
\cite{note_CP_scaling} that $w_c^{(3)}\of{\rho\to 0}\to 0$, there 
are no general theoretical estimates concerning which values
$w_c^{(3)}\neq 0$ are attained for $\rho\neq 0$. This lack of knowledge 
is overcome in the main text by resorting to the results obtained from 
the effective approach. 
Still this leaves open the issue, to which extent the non-universal amplitude
$w_0^{(3)}$ depends on the kind of colloids used. 
We recall that within our analysis the dependence on the kind of solvent 
used is taken into account by adopting the corresponding values for the 
non-universal parameters. Concerning the kind of colloids, at least 
two main influences of the colloids on the solvent, i.e., the excluded 
volume and the strong adsorption of one of the components, are taken into
account via $R$ and the BCs for the solvent OP at the colloid surfaces 
(i.e., $\hs=\infty$ in \eref{eq:lgh}), respectively.
The non-universal amplitude may depend on the direct colloid-solvent
interaction, e.g., via a non-universal amplitude related to a finite 
surface field $\hs<\infty$. The direct interactions between the colloids 
can be expected to enter the non-universal amplitude, too. Softly 
repulsive colloid-colloid interactions may be taken into account 
by using an effective HS diameter $\sigma$ (see, e.g.,
\eref{eq:eff_hs_diameter}) instead of $2R$ (which, in turn corresponds 
to a non-universal amplitude $w_0^{(3)}$ proportional to $\sigma/\of{2R}$).



\end{thebibliography}
\end{document}